\documentclass[journal,10pt]{IEEEtran}
\usepackage{graphicx}
\usepackage{multirow}
\usepackage{caption}
\usepackage{subcaption}
\usepackage{mathtools}
\usepackage{bm, amsmath, amssymb}
\usepackage{amsmath}
\usepackage{amsfonts}
\usepackage {amssymb,amsmath}
\usepackage{cite}
\usepackage{epstopdf}
\usepackage{microtype}
\usepackage{color}
\DeclareGraphicsExtensions{.pdf,.png,.jpg}

\newcommand {\Define} {\stackrel {\Delta} {=}  }
\newcommand{\mya}{\mathrel{\overset{\makebox[0pt]{{\tiny(a)}}}{=}}}
\newcommand{\myapproxa}{\mathrel{\overset{\makebox[0pt]{{\tiny(a)}}}{\approx}}}
\newcommand{\myb}{\mathrel{\overset{\makebox[0pt]{{\tiny(b)}}}{=}}}
\newcommand{\myapproxb}{\mathrel{\overset{\makebox[0pt]{{\tiny(b)}}}{\approx}}}
\newcommand{\myc}{\mathrel{\overset{\makebox[0pt]{{\tiny(c)}}}{=}}}
\newcommand{\myapproxc}{\mathrel{\overset{\makebox[0pt]{{\tiny(c)}}}{\approx}}}

\newtheorem{theorem}{Theorem}
\newtheorem{corollary}{Corollary}

\usepackage{cite}

\DeclareMathOperator{\sinc}{sinc}

\begin{document}
\title{Low Complexity Precoding and Detection in Multi-user Massive MIMO OTFS Downlink}
\author{\IEEEauthorblockN{Brijesh Chander Pandey, Saif Khan Mohammed, P. Raviteja, Yi Hong and Emanuele Viterbo}
\IEEEauthorblockA{ \thanks{Brijesh Chander Pandey is with the Department of Electrical Engineering, Indian Institute of Technology Delhi and Saif Khan Mohammed is with the Department of Electrical Engineering, Indian Institute of Technology Delhi (IITD), New Delhi, India. Email: saifkmohammed@gmail.com. This work is an outcome of the Research and Development work undertaken in the project under the Visvesvaraya PhD Scheme of Ministry of Electronics and Information Technology, Government of India, being implemented by Digital India Corporation (formerly Media Lab Asia) and the multi-institute project on an ``Indigineous 5G Test Bed''. This work was also supported by the EMR funding from the Science and Engineering
Research Board (SERB), Department of Science and Technology (DST),
Government of India. P. Raviteja (raviteja.patchava@monash.edu), Yi Hong (yi.hong@monash.edu) and Emanuele Viterbo (emanuele.viterbo@monash.edu) are with the Department of Electrical and Computer Systems Engineering, Monash University, Australia.}}
}
\maketitle

\vspace{-14mm}
\begin{abstract}
We consider the problem of degradation in performance of multi-carrier multi-user massive MIMO systems when channel induced Doppler spread is high. Recently, Orthogonal Time Frequency Space (OTFS) modulation has been shown to be robust to channel induced Doppler spread. In OTFS based systems, information symbols are embedded in the delay-Doppler (DD) domain where they are {\em jointly} modulated to generate the time-domain transmit signal. Due to the multi-path delay and Doppler shifts, the DD domain information symbols need to be jointly demodulated at the receiver.
For multi-carrier based communication (e.g., Orthogonal Frequency Division Multiplexing (OFDM)), massive MIMO systems have been shown to achieve high spectral and energy efficiency with low complexity multi-user precoding in the downlink. Extending the same to OTFS based downlink multi-user massive MIMO systems is challenging due to the requirement for joint demodulation of all information symbols at the user terminal (UT). In this paper, we solve this problem by proposing a novel OTFS based multi-user precoder at the base station (BS) and a corresponding low complexity detector (LCD) at the user terminals (UTs), which allows for separate demodulation of each DD domain information symbol at the UT. The complexity of the proposed precoder increases only linearly with increasing number BS antennas $Q$ and the number of UTs. We show, through analysis, that as $Q$ increases (with total transmitted power decreased linearly with $Q$), the proposed low complexity detector achieves a sum spectral efficiency close to that achieved with optimal joint demodulation at each UT. Numerical simulations confirm our analysis and also show that the spectral efficiency and error rate performance of the proposed OTFS based massive MIMO precoder (with the proposed LCD detector at each UT) is significantly more robust to channel induced Doppler spread when compared to OFDM based multi-user massive MIMO systems.                       
\end{abstract}

\begin{IEEEkeywords}
	OTFS, Massive MIMO, Doppler Spread, Multi-user, Precoder.
\end{IEEEkeywords}
\section{Introduction}
In this paper we consider low complexity precoding and detection for downlink massive MIMO systems where information is embedded
in the Delay-Doppler (DD) domain. Recently, Orthogonal Time Frequency Space (OTFS) modulation has been introduced, where information
is embedded in the DD domain \cite{HadaniOTFS1, HadaniOTFS2, HadaniOTFS3}.  
This improves robustness towards channel induced Doppler spread which is otherwise known to severely degrade
the performance of multi-carrier systems (e.g., Orthogonal Frequency Division Multiplexing (OFDM)) where information is embedded in the Time-Frequency (TF)
domain \cite{HadaniOTFS1}, \cite{SKM4}.

In OTFS modulation, DD domain information symbols interfere with each other due to the channel induced delay and Doppler shifts, and therefore
they need to be jointly demodulated at the receiver. This can be challenging in the downlink where the UTs are receivers.
Towards this end, \cite{channel, Emanuele2, lmpa}, \cite{OTFSMP} have considered low-complexity message passing (MP) based detectors which have good error rate performance when the DD domain channel is sparse i.e., small number of channel paths.
However, in rich scattering environments with a large number of channel multi-paths, the performance of MP based detector degrades. In \cite{MMSEOTFS} a Linear Minimum Mean Squared Error (LMMSE) estimator based OTFS detector has been proposed whose complexity increases
with increasing delay and Doppler spread. In \cite{mimootfs}, a point-to-point MIMO-OTFS system
has been considered for high Doppler spread channels, for which a Markov Chain Monte Carlo (MCMC) sampling based detector has been proposed. However, the complexity of this MCMC based detector is $O(M^3 N^3)$ where $M$ and $N$ are the number of sub-divisions of the delay and Doppler domain respectively. The complexity of this MCMC detector can be prohibitive for large $(M,N)$.
A low-complexity Linear MMSE detector has also been proposed in \cite{LMMSEOTFS} having complexity $O(M^2N^2P)$ ($P$ is the number of
channel multi-paths), though this complexity can still be high for OTFS systems where $(M,N)$ is large. In \cite{OTFSBayes},
low complexity OTFS detection based on a variational Bayes approach is proposed having overall complexity $O(MNP)+O(MN \log(MN))$.  

Next generation communication systems are expected to achieve higher throughput and reliability at much higher mobile speed when compared to Fourth Generation systems \cite{IMT2020}.
As next generation systems are expected to operate in higher frequency bands, achieving high throughput at high mobile speed is a challenge due to increased channel induced Doppler spread.
Massive MIMO systems based on large antenna arrays at the base station (BS) have been proposed
for improving the spectral and energy efficiency of multi-user cellular wireless communication systems \cite{ErikNextGen}.
However, to the best of our knowledge there is not much work done so far on {\em large antenna array} based {\em downlink precoding} for {\em OTFS multi-user} systems.

Recently in \cite{Yaru2020}, the authors have proposed a large-antenna array based low-complexity receiver for decoding the downlink OTFS signal transmitted from a single-antenna BS. However, the authors ignore inter-carrier interference (ICI) and inter-symbol interference (ISI) as they assume ideal pulse shaping transmit and receiver waveforms which satisfy the bi-orthogonality condition (see equation $(17)$ in \cite{Yaru2020}). Such ideal pulses however do not exist. Further, the Doppler shift of each channel path is assumed to be an integer multiple of the Doppler resolution in the DD domain, and therefore interference due to fractional Doppler has also been ignored in \cite{Yaru2020}. Also, in \cite{Yaru2020}, only a single user scenario is considered with no precoding at the BS as only single-antenna BS transmission is considered.
In \cite{ChEstOTFS3}, a path division multiple access OTFS massive MIMO system has been proposed where the angle domain is used to separate
the uplink signal from different UTs. This method however assumes precise knowledge of the antenna steering vector at the BS and also accurate
knowledge of the angle of arrivals from the UTs.  

In \cite{Zemen2019}, OTFS based single user beamforming is proposed. The OTFS precoder transforms the information symbols into TF domain symbols on which beamforming is performed.  
Due to the OTFS precoder, each information symbol is spread over the entire TF domain and therefore at the receiver,
a linear MMSE estimator based interference cancellation detector is used. For each information symbol, the complexity of
the linear MMSE based detection is at least $O(MN)$. Since there are $MN$
information symbols in an OTFS frame, the total detection complexity is at least $O(M^2 N^2)$.  
          
In this paper, we consider a {\em multi-user} massive MIMO downlink OTFS system where a BS having a {\em large} uniform rectangular antenna array (URA) with $Q= Q_h Q_v$ antennas ($Q_h$ rows and $Q_v$ columns) serves $K$ user terminals (UTs) each having a single antenna.
We consider practical non-ideal pulse shaping and fractional Doppler which results in ICI, ISI and Doppler interference, due to which the effective DD domain channel matrices between the BS and the UTs
are not as sparse as that when ideal pulses and integral Doppler is assumed.
In such a scenario, multi-antenna information precoding in the DD domain can have very high complexity since all $KMN$ information symbols have to be jointly precoded and transmitted from the $Q$ BS antennas. There is therefore a need to design {\em low complexity} multi-user precoding and detection methods for massive MIMO OTFS systems.  Specific contributions of our paper are:
\begin{enumerate}
\item With rectangular pulse shaping and fractional Doppler, in Section \ref{secbeamformer} we propose a low-complexity DD domain based multi-user precoder for massive MIMO OTFS systems where perfect channel state information (CSI) of all UTs is available at the BS. The precoding complexity is
$O\left(QMN^2 \sum\limits_{s=1}^K L_s \right)$ where $L_s$ is the number of channel paths between the BS and the $s$-th UT. Note that the complexity increases only {\em linearly} with the number of BS antennas $Q$ and the number of UTs $K$. 
\item With this proposed multi-user precoder, the optimal detector at each UT has prohibitive complexity as it jointly decodes all the $MN$ DD domain information symbols in an OTFS frame. Therefore, in Section \ref{seclcd} we propose a low complexity DD domain detector (LCD) at the UT, which performs {\em separate} detection for each information symbol and therefore has lower complexity than the optimal detector. Due to separate detection of each information symbol, the LCD detector
along with the TF to DD domain converter has only a $O(MN \log(MN))$ complexity, which is independent of the amount of delay/Doppler spread and the number of channel paths.  
\item In Section \ref{seclcd} we also derive an expression for the achievable sum spectral efficiency (SE) with the proposed low complexity detector. Analysis of this expression reveals that
as the number of BS antennas becomes large (with total transmitted power reduced linearly with increasing $Q$), the sum SE achieved by the proposed LCD detector is close to that achieved by the optimal detector.   
\item Analysis reveals that in the large antenna regime, the SE of the proposed LCD detector is limited only by AWGN and is almost independent of multi-user and inter-symbol interference. This is similar to the effect of vanishing multi-user and inter-symbol interference in Time-Frequency domain based massive MIMO systems where linear precoders are used \cite{MassiveMIMOBook}.      
\item Numerical simulations confirm that, in the large antenna regime, the proposed LCD detector achieves sum SE close to that achieved by the optimal detector. Through numerical simulations, we also compare the sum SE achieved by the proposed multi-user precoder and LCD detector with that achieved by OFDM massive MIMO systems based on Maximum Ratio Transmission (MRT) multi-user precoding \cite{MassiveMIMOBook}. These simulations reveal that the proposed DD domain based multi-user massive MIMO precoding and LCD detection achieves {\em significantly better} SE than OFDM massive MIMO systems in channels where Doppler shift is high.    
\item In Section \ref{errorrateperf} we compare the coded error rate performance of OTFS and OFDM downlink massive MIMO
systems when the BS has {\em imperfect} channel state information (CSI). Simulations reveal that indeed the proposed low-complexity OTFS precoder with the LCD detector achieves significantly better coded error rate performance than OFDM massive MIMO with much less channel estimation overhead.         
\end{enumerate}
{\em Notations:} For integers $x$ and $M$, $[ x ]_{M}$ denotes the smallest non-negative integer congruent to $x$ modulo $M$. For any real $y$, $\lfloor y \rfloor$ is the greatest integer
less than or equal to $y$. The zero mean circular symmetric complex Gaussian distribution having variance $N_0$ is denoted by
${\mathcal C}{\mathcal N}(0, N_0)$. For any square matrix ${\bf A}$, $\mbox{Tr}({\bf A})$ and $\left\vert {\bf A} \right\vert$ denote the trace and determinant of ${\bf A}$, respectively.
$\mbox{\small{Diag}}(a_1, a_2, \cdots, a_n)$ denotes a square $n \times n$ diagonal matrix whose element in the $i$-th row and $i$-th column is $a_i, i=1,2,\cdots,n$.   

\section{System Model}
\label{sysmodelsec}
We consider the downlink of an OTFS based multi-user massive MIMO system with $Q$ co-located antennas at the base station (BS) and a single antenna at each of the $K$ user terminals (UTs). The antenna array at the BS is in the form of a Uniform Rectangular Array (URA) with distance $d$ between adjacent antenna elements. The number of rows and columns of antenna elements in the URA are $Q_v$ and $Q_h$ respectively, i.e., $Q = Q_h Q_v$. The Delay-Doppler channel between the $q$-th BS antenna and the $s$-th UT is given by \cite{Jakes, Bello}

{\vspace{-5mm}
\small
\begin{eqnarray}
\label{eqn1}
h_{q,s}(\tau,\nu)   =   \sum\limits_{i=1}^{L_s} \, h_{q,s,i} \, \delta(\tau - \tau_{s,i}) \,  \delta(\nu - \nu_{s,i}) \,,  \nonumber \\
 s=1,2, \cdots, K \,\,,\,\, q=1,2,\cdots, Q
\end{eqnarray}    
\normalsize}
where $L_s$ denotes the number of channel paths between the BS and the $s$-th UT. Further, $\tau_{s,i}$ and $\nu_{s,i}$ are the delay and Doppler along the $i$-th path $(i = 1,2, \cdots, L_s)$ between the BS and the $s$-th UT.
In (\ref{eqn1}), $\delta(\cdot)$ is the impulse function. Further, $h_{q,s,i}$ is the complex channel gain between the $q$-th BS antenna and the $s$-th UT along the $i$-th path and is given by \cite{URAPaper}
\begin{eqnarray}
\label{eqn3}
h_{q,s,i}  =  g_{s,i} \, e^{j \frac{2 \pi d}{\lambda}\left( [ q-1]_{_{Q_h}} \sin \phi_{s,i} \sin \theta_{s,i}+ \left\lfloor  \frac{q-1}{Q_h} \right\rfloor \cos \theta_{s,i}\right)}\,\,, \nonumber \\
q=1,2, \cdots,Q \,,\, s = 1,2,\cdots, K \,,\, i=1,2,\cdots, L_s
\end{eqnarray}
where $g_{s,i}$ models the complex channel gain of the $i$-{th} path, i.e., $g_{s,i}$ are i.i.d. $ \mathcal{C} \mathcal{N}(0, \beta_{s,i})$, $ s =1,2,\cdots,K \,,\, i=1,2,\cdots,L_s $.
Further, let the vector of channel path gains between the $s$-th UT and the BS be denoted by
\begin{eqnarray}
\label{gsvecdef}
{\bf g}_s & \Define & \left( \, g_{s,1} \,,\, g_{s,2} \,,\, \cdots \,,\, g_{s,L_s} \, \right)^T \,,\, s=1,2,\cdots,K.
\end{eqnarray}
In (\ref{eqn3}), $\theta_{s,i}$ and $\phi_{s,i}$ are respectively the zenith and azimuthal angles of departure for the $i$-th channel path from the BS to the $s$-th UT.\footnote{\footnotesize{We assume that for any two channel paths to a UT, both the corresponding zenith and azimuthal angles are not equal. That is, for all $s =1,2,\cdots, K$ and any $i \ne k$, $(\theta_{s,i}, \phi_{s,i}) \ne (\theta_{s,k},\phi_{s,k})$, i.e., $\theta_{s,i} = \theta_{s,k}$ and $\phi_{s,i} = \phi_{s,k}$ if and only if $i = k$.  In practical scenarios, due to randomly distributed scatterers, two UTs may have the same zenith and azimuthal angles for some paths between the BS and these UTs. However, in this paper we consider that the BS schedules any two such UTs onto different physical resource, in order that they do not interfere with each other. Therefore, in this paper we propose joint DD domain precoding only for those UTs for which, the zenith and the azimuthal angles for any channel paths to these UTs are not both equal, i.e., for any $s \ne s'$ and all $i = 1,2,\cdots, L_s$, $k=1,2,\cdots,L_{s'}$, $(\theta_{s,i}, \phi_{s,i}) \ne (\theta_{s',k},\phi_{s',k})$.}} Also, $\lambda$ is the wavelength of the carrier.

In this paper we consider coding across several OTFS frames which are sequential in time.
The zenith and azimuthal angles of departure, the Doppler shift, delay and channel gain for each path are assumed to be fixed for the entire duration of a codeword. As an example, with OTFS frames of duration $0.25$ milliseconds (ms) and a codeword spanning hundred OTFS frames, the total duration of the codeword is $25$ ms. With a bandwidth of $5$ MHz, the resolution of the delay domain is $1/(5 \mbox{\tiny{MHz}}) = 0.2 \mu s$. The delay of a channel path can change by an order of $0.2 \mu s$ if and only if the UT travels a distance of roughly $60$ m (i.e., distance travelled in $0.2 \mu s$ at speed of light). Even for a UT with a mobile speed of $360$ Km/hr (i.e., $100$ m/s), it would take $600$ ms for this to happen, which is much greater than the $25$ ms duration of a codeword.
In this paper, we therefore assume the zenith and azimuthal angles of departure, the Doppler shift, delay and channel gain for each path to be fixed for a few tens of ms, i.e., the entire duration of the codeword. Subsequently, these parameters which are fixed for the entire codeword duration, are denoted by
\begin{eqnarray}
\label{paramdefeqn}
{\mathcal P} \Define \left\{ \{L_s\}_{s=1}^K, \{\theta_{s,i}, \phi_{s,i}, \tau_{s,i}, \nu_{s,i}, g_{s,i} \}_{s=1, i=1}^{K, L_s} \right\}.
\end{eqnarray}  
Let $x_q(t)$ denote the time-domain signal transmitted from the $q$-th BS antenna.     
The time-domain signal received at the $s$-th UT is given by \cite{Bello}

\begin{figure}[t]
\vspace{-0.5 cm}
\centering
\includegraphics[width= 2.8 in, height= 1.9 in]{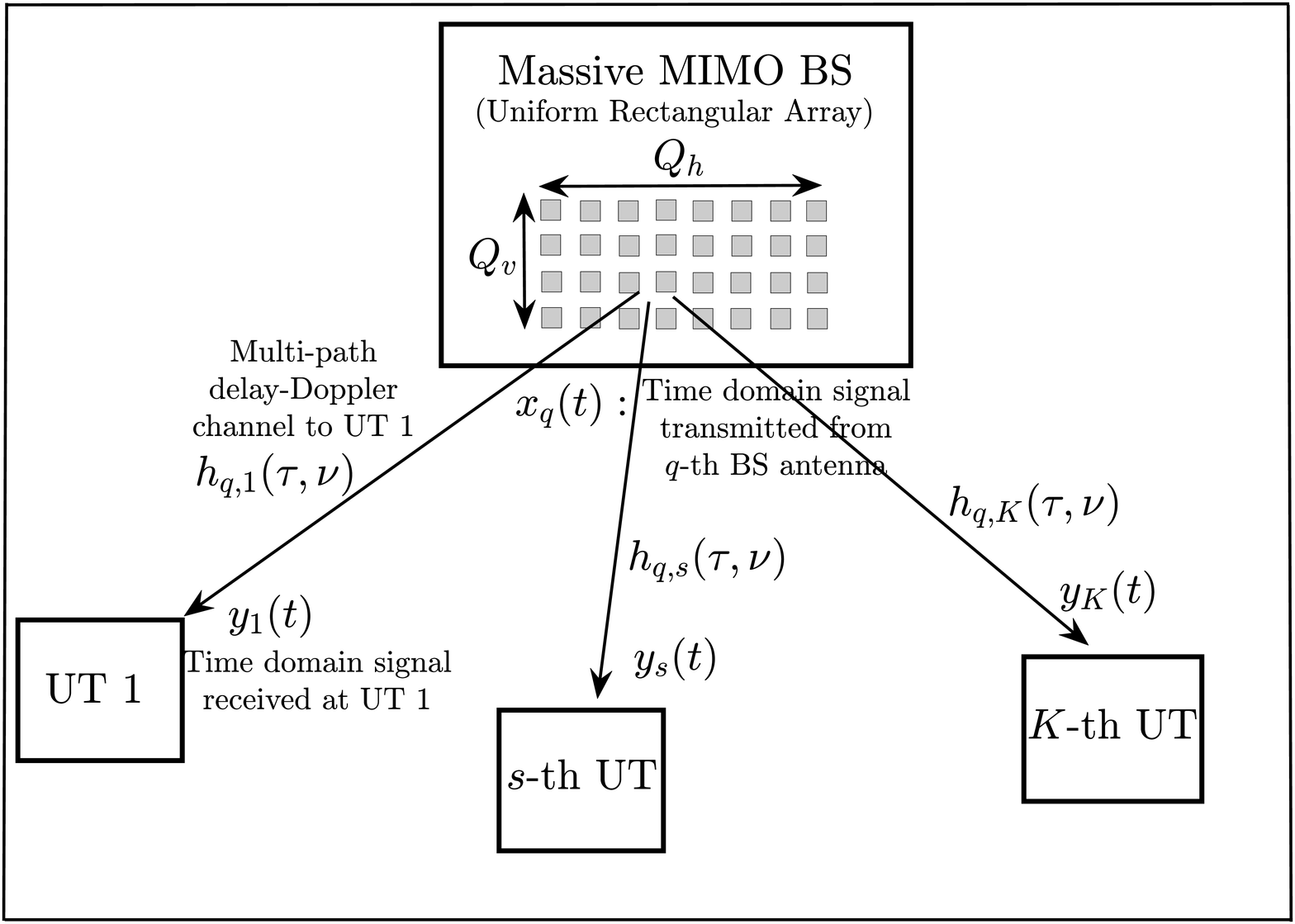}
\vspace{-0.1 cm}
\caption{A multi-user massive MIMO downlink system.} 
\vspace{-0.1cm}
\label{sm_fig}
\end{figure}

{\vspace{-6mm}
\small
\begin{eqnarray}
\label{eqn4}
y_s(t) & \hspace{-3mm}  = &  \hspace{-3mm} \sum\limits_{q=1}^Q \int \int h_{q,s} (\tau,\nu)  x_q(t - \tau)  e^{j 2 \pi \nu (t - \tau)}  d\tau \, d\nu \,+\,  {\Tilde w_s(t)}  \nonumber \\
& \hspace{-3mm}  \mya &   \hspace{-3mm} \sum\limits_{q=1}^Q \sum\limits_{i=1}^{L_s}  h_{q,s,i} \, x_q(t - \tau_{s,i}) \,  e^{j 2 \pi  \nu_{s,i} (t - \tau_{s,i})} \, + \, {\Tilde w_s}(t)
\end{eqnarray}
\normalsize}
where ${\Tilde w_s}(t)$ is the additive white Gaussian noise (AWGN) at the $s$-th UT and step (a) follows from (\ref{eqn1}).
The multi-user massive MIMO downlink system is depicted in Fig.~\ref{sm_fig}. 
\section{OTFS Modulation and Demodulation}
The DD domain is $T$ seconds wide along the delay domain and $\Delta f = 1/T$ Hz wide along the Doppler domain. The delay domain is further sub-divided
into $M$ equal sub-divisions each $T/M$ seconds wide. Similarly the Doppler domain is sub-divided into $N$ equal sub-divisions each $\Delta f/N$ Hz wide.
The combination of a sub-division along the delay domain and a sub-division along the Doppler domain is referred to as a Delay Doppler Resource Element (DDRE).
There are therefore $MN$ DDREs in one OTFS frame and each DDRE carries one information symbol. Each DDRE is denoted by its index $k = 0,1,\cdots, N-1$ along the
Doppler domain and its index $l=0,1,\cdots, M-1$ along the delay domain.

%In OFDM systems, channel induced Doppler shift spreads
%interference from the information transmitted on a particular sub-carrier to several neighbouring sub-carriers.
%However, in OTFS, channel induced Doppler shift only shifts the
%location of the information symbol from its transmit DDRB location to another DDRB at the receiver, but does not spread as much interference to other DDRBs. For example,
%let us consider a channel consisting of two paths, with the first path inducing a delay of $5T/M$ and a Doppler shift of $2 \Delta f/N$ while the second path induces a delay of $3T/M$ and a Doppler shift of $- 3 \Delta f/N$. An information symbol transmitted on the $(k,l)$-th DDRB would then be received on two different DDRBs at the receiver, namely the $([k+2]_N, [l+5]_M)$-th DDRB due to the first path and the $([k-3]_N, [l+3]_M)$-th DDRB due to the second path (here $[x]_M$ denotes the smallest non-negative integer which is congruent to $x$ modulo $M$). These multiple received symbols can then be combined coherently to achieve improved robustness towards channel induced Doppler shift. 
%In \cite{HadaniOTFS1, otfsmmwave}, the performance of a single-antenna based OTFS system has been studied and its reliability performance is reported to be better than that of an OFDM based system.

At the $q$-th BS antenna, the DD domain transmit symbols $x_q[k,l], k=0,1,\cdots,N-1, l=0,1,\cdots, M-1$ are converted to the time-domain signal $x_q(t)$ using OTFS modulation \cite{HadaniOTFS1}. At the UT, the received time-domain signal $y(t)$ is transformed back to the DD domain using OTFS demodulation. For OTFS modulation, the DD domain signal $x_q[k,l]$ is first transformed to the Time-Frequency (TF) domain signal $X_q[n,m]$ using the Inverse Symplectic Finite Fourier Transform (ISFFT), i.e. 

{\vspace{-4mm}
\small
\begin{eqnarray}
\label{eqn5}
X_q[n,m] & = & \frac{1}{\sqrt{MN}} \sum\limits_{k=0}^{N-1}\sum\limits_{l=0}^{M-1} x_q[k,l] \, e^{j 2 \pi \left( \frac{n k}{N} - \frac{m l}{M}  \right)}  \nonumber \\
& & \hspace{-7mm} n=0,1,\cdots,N-1 \,,\, m=0,1,\cdots, M-1.
\end{eqnarray}
\normalsize}
Next, $X_q[n,m]$ is then transformed to the time-domain signal ${\Tilde x}_q(t)$ using the Heisenberg transform\footnote{\footnotesize{Transformation of the TF signal $X_q[n,m]$ to the time-domain can also be performed using the Inverse Fourier Transform as considered in \cite{ChEstOTFS1, ChEstOTFS2, ChEstOTFS3}. However, the use of the Inverse Fourier Transform requires a Cyclic-Prefix (CP) for each $n=0,1,\cdots, N-1$, i.e., $N$ CPs in each OTFS frame. In \cite{PracPulseOTFS} it has been shown that using the Heisenberg transformation (instead of the Inverse Fourier Transform)
requires only {\em one} CP for the entire OTFS frame without any performance penalty, i.e., the CP overhead is significantly less when using the Heisenberg transformation.}} 

{\vspace{-5mm}
\small
\begin{eqnarray}
\label{eqn6pr}
{\Tilde x}_q(t) & \hspace{-3mm}  = & \hspace{-3mm} \sum\limits_{n=0}^{N-1} \sum\limits_{m=0}^{M-1} X_q[n,m]  \, g_{tx}(t - nT) \, e^{j 2 \pi m \Delta f (t - nT)}
\end{eqnarray}   
\normalsize}    
where the transmit pulse $g_{tx}(\cdot)$ is rectangular and is given by
\begin{eqnarray}
\label{gtxpulse_eqn}
g_{tx}(t) \Define \left\{
  \begin{array}{@{}ll@{}}
   \frac{1}{\sqrt{T}} \, &, \, \mbox{if} \,\,\, 0 \leq t < T \\
   0  \, & \,,\, \mbox{otherwise}.
  \end{array}\right.
\end{eqnarray}
A CP of duration equal to the maximum channel path delay, i.e.\footnote{\footnotesize{The CP helps in avoiding interference between
OTFS frames transmitted one after the other in the time-domain.}}

{\vspace{-6mm}
\small
\begin{eqnarray}
\tau_{max}  & \Define &  \max\limits_{s=1,2,\cdots,K \atop i=1,2,\cdots, L_s}  \tau_{s,i}
\end{eqnarray} 
\normalsize}
is appended to ${\Tilde x}_q(t) $ resulting in the time-domain signal $x_q(t)$ to be transmitted from the $q$-th BS antenna, i.e.

{\vspace{-4mm}
\small
\begin{eqnarray}
\label{eqn6}
x_q(t) & \hspace{-3mm} = \begin{cases}
\sum\limits_{n=-1}^{N-1} \sum\limits_{m=0}^{M-1} X_q[[n]_{_N},m] g_{_{tx}}(t - nT) \, e^{j 2 \pi m \Delta f (t - nT)}, \hspace{-3mm} \\
 \hspace{43mm}  -\tau_{max} \leq t <  NT  \hspace{-3mm} \\
 0 \,,\, t <  - \tau_{max} \,\,\mbox{\small{and}} \,\, t  \geq NT \, .
\end{cases}
\end{eqnarray}
\normalsize}
In (\ref{eqn6}), the TF symbol $X_q[n,m], n \geq 0$ contributes to the overall transmit signal $x_q(t)$ only through the term $X_q[n,m] g_{tx}(t - nT) \, e^{j 2 \pi m \Delta f (t - nT)}$
and therefore since $g_{tx}(\cdot)$ is time-limited to the interval $[0 \,,\, T)$, it follows that this term is limited to the interval $[nT \,,\, (n+1)T)$ along the time-domain and roughly along the interval $[m \Delta f \,,\, (m+1)\Delta f]$ along the frequency domain ($[nT \,,\, (n+1)T) \times [m \Delta f \,,\, (m+1)\Delta f]$ is referred to as the $(m,n)$-th time-frequency resource element (TFRE)).
In the R.H.S. of (\ref{eqn6}), the summation term due to $n=-1$ corresponds to the CP and we limit it to the time-interval $[ - \tau_{max}, 0 )$. Since $m=0,1,\cdots, (M-1)$ and $n=-1,0,1,\cdots, (N-1)$, the transmit signal $x_q(t)$ is therefore time-limited to the interval $[-\tau_{max} \,,\, NT)$ and has a bandwidth of $M \Delta f$ Hz.
For OTFS demodulation at the $s$-th UT, the received time-domain signal $y_s(t)$ is first transformed to the TF domain using the Wigner transform \cite{HadaniOTFS1} i.e.

{\vspace{-5mm}
\small
\begin{eqnarray}
\label{eqn7}
Y_s[n,m] & = & \int\limits_{0}^{NT} g_{rx}^*(t - nT) \, y_s(t) \, e^{-j 2 \pi m \Delta f (t - nT)} \, dt \nonumber \\
& & \hspace{-7mm} n=0,1,\cdots,N-1 \,,\, m=0,1,\cdots, M-1
\end{eqnarray}
\normalsize
}
where $g_{rx}(\cdot)$ is the receive pulse which we take to be rectangular i.e., $g_{rx}(t) = g_{tx}(t)$.
The TF domain signal is then transformed back to the DD domain using SFFT i.e.,

{\vspace{-4mm}
\small
\begin{eqnarray}
\label{eqn8}
{\widehat x_s}[k,l] & = &  \frac{1}{\sqrt{MN}} \sum\limits_{n=0}^{N-1}\sum\limits_{m=0}^{M-1}  Y_s[n,m] \, e^{-j 2 \pi \left( \frac{n k}{N} - \frac{m l}{M}  \right)}  \nonumber \\
& & \hspace{-7mm} k=0,1,\cdots,N-1 \,,\, l=0,1,\cdots, M-1.
\end{eqnarray}
\normalsize}
Let ${\widehat {\bf x}_s} \in {\mathbb C}^{MN \times 1}$ denote the vector of these received DD domain symbols at the $s$-th UT as given by (\ref{eqn9}) (see top of next page).
\begin{figure*}
{\vspace{-9mm}
\small
\begin{eqnarray}
\label{eqn9}
{\widehat {\bf x}_s} &  \Define  &  \left(  {\widehat x_s}[0,0] , \cdots, {\widehat x_s}[0,M-1], {\widehat x_s}[1,0], \cdots, {\widehat x_s}[1,M-1], \cdots, {\widehat x_s}[N-1,0], \cdots,    {\widehat x_s}[N-1,M-1]\right)^T, \nonumber \\
{{\bf x}_q} &  \Define  &  \left(  { x_q}[0,0] , \cdots, { x_q}[0,M-1], {x_q}[1,0], \cdots, { x_q}[1,M-1], \cdots, { x_q}[N-1,0], \cdots,    {x_q}[N-1,M-1]\right)^T, \nonumber \\
{{\bf w}_s} &  \Define  &  \left(  { w_s}[0,0] , \cdots, { w_s}[0,M-1], {w_s}[1,0], \cdots, { w_s}[1,M-1], \cdots, { w_s}[N-1,0], \cdots,    {w_s}[N-1,M-1]\right)^T. 
\end{eqnarray}
\normalsize
}
\end{figure*}
From Appendix \ref{appendixRavi} in this paper and \cite{channel} it follows that\footnote{\footnotesize{Note that, just as in \cite{channel}, we consider the path delays $\tau_{s,i}, i=1,2,\cdots,L_s$
to be integer multiples of $T/M =1/(M \Delta f)$ since the sampling time duration $1/(M \Delta f)$ is small for wide bandwidth systems and therefore
it suffices to approximate the path delay with its nearest integer multiple of $1/(M \Delta f)$ \cite{DTse}.
Therefore, $ l_{\tau_{s,i}} \Define  \tau_{s,i} M \Delta f$ is an integer.}}

{\vspace{-6mm}
\small
\begin{eqnarray}
\label{eqn10}
{\widehat {\bf x}_s} & = &  \sum\limits_{q=1}^Q {\bf H}_{q,s} \, {\bf x}_q \, + \, {\bf w}_s
\end{eqnarray}
\normalsize}
where ${\bf x}_q \in {\mathbb C}^{MN \times 1}$ is the vector of DD domain symbols to be transmitted from the $q$-th BS antenna, and is given by (\ref{eqn9}).
In (\ref{eqn10}), ${\bf w}_s$
is the vector of i.i.d. ${\mathcal C}{\mathcal N}(0,N_0)$ additive noise samples in the DD domain and is given by (\ref{eqn9}). In (\ref{eqn9}), $w_s[k,l]$ is the received noise sample
at the $s$-th UT in the $(k,l)$-th DDRE and is given by (\ref{noisekleqn}) in Appendix \ref{appendixRavi}.
The matrix ${\bf H}_{q,s} \in {\mathbb C}^{MN \times MN}$
is the effective DD domain channel between the $s$-th UT and the $q$-th BS antenna and is given by

{\vspace{-5mm}
\small
\begin{eqnarray}
\label{eq_channel}
{\bf H}_{q,s} \Define \sum\limits_{i=1}^{L_s} h_{q,s,i}  \, {\bf A}_{s,i}
\end{eqnarray}  
\normalsize}
where ${\bf A}_{s,i} \in {\mathbb C}^{MN \times MN}$ and its entry in the $(kM+l+1)$-th row and $(k^{\prime}M+l^{\prime}+1)$-th column is given by $a_{s,i,k,l}[k^{\prime},l^{\prime}]$ in (\ref{asikleqn})
of Appendix \ref{appendixRavi}. 
\begin{figure*}
{\vspace{-8mm}
\begin{eqnarray}
\label{eqn13}
{{\bf u}_s} &  \Define  &  \left(  { u_s}[0,0] , \cdots, { u_s}[0,M-1], {u_s}[1,0], \cdots, { u_s}[1,M-1], \cdots, { u_s}[N-1,0], \cdots,    {u_s}[N-1,M-1]\right)^T.
\end{eqnarray}
\begin{eqnarray*}
\hline
\end{eqnarray*}
\vspace{-3mm}
}
\end{figure*}

\section{Proposed Multi-user Downlink Precoding in DD Domain}
\label{secbeamformer}
Let $u_s[k,l] \sim {\mathcal C}{\mathcal N}(0,1) \,,\, k=0,1,\cdots,N-1, l=0,1,\cdots,M-1$ denote the i.i.d. DD domain information symbols
to be communicated to the $s$-th UT on the $(k,l)$-th DDRE. Also, let ${\bf u}_s \in {\mathbb C}^{MN \times 1}$
denote the vector of all $MN$ information symbols for the $s$-th UT, which is given by (\ref{eqn13}) (see top of next page).
Assuming perfect knowledge of the effective DD domain channel matrices ${\bf H}_{q,s} \,,\, q=1,2,\cdots,Q \,,\, s=1,2,\cdots, K$ at the BS,
in the following we propose to precode ${\bf u}_s$ into the DD domain transmit signal vectors ${\bf x}_q \,,\, q=1,2,\cdots, Q$, i.e.

\begin{figure}[t]
\vspace{-0.8 cm}
\centering
\includegraphics[width= 3.3 in, height= 2.2 in]{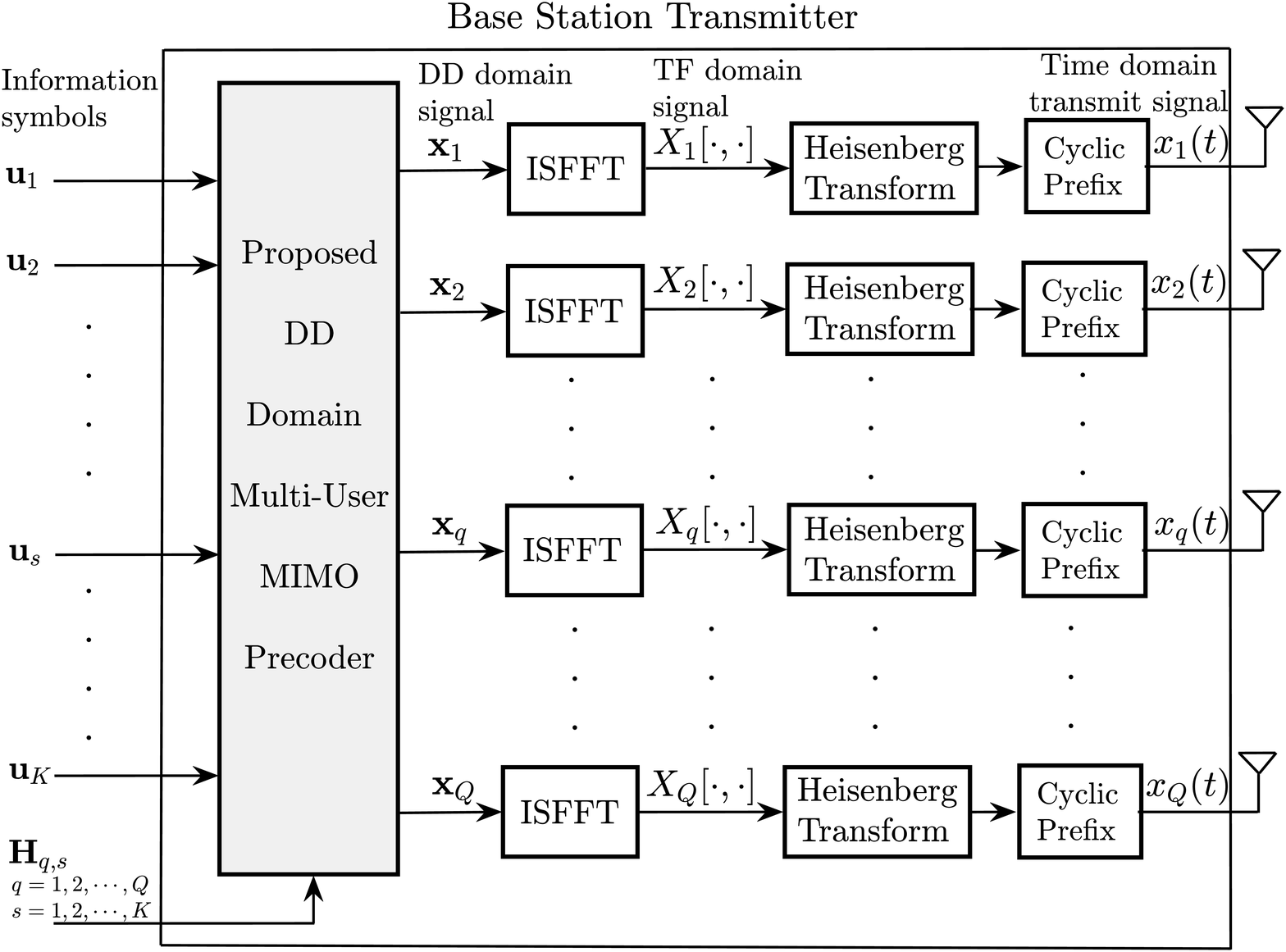}
\vspace{-0.1 cm}
\caption{Block diagram of the BS downlink transmitter with the proposed OTFS Multi-user Massive MIMO Precoder.} 
\vspace{-0.13cm}
\label{tx_fig}
\end{figure}

{\vspace{-4mm}
\small
\begin{eqnarray}
\label{eqn14}
{\bf x}_q   \,  =  \,  \sqrt{\frac{E_T}{\eta}} \, \sum\limits_{s =1}^K \, {\bf H}^H_{q,s}\, {\bf u}_s \,\,\,,\,\,\, \eta \Define  Q M N \sum\limits_{s=1}^K  \sum\limits_{i=1}^{L_s} \beta_{s,i}.
\end{eqnarray}
\normalsize}
The proposed precoder in (\ref{eqn14}) incorporates spatial precoding as the DD domain signal ${\bf x}_q$
transmitted from the $q$-th BS antenna depends on the DD domain channel matrices ${\bf H}_{q,s}, s=1,2,\cdots, K$
between the UTs and the $q$-th BS antenna, and further from (\ref{eq_channel}) we know that ${\bf H}_{q,s}$ depends on $h_{q,s,i}$
which in turn depends on the zenith and azimuthal angles of departure between the BS and the $s$-th UT. Later in Section \ref{seclcd}, we
show that with a large BS antenna array, this precoder allows for separate detection of each DD domain
information symbol (i.e., low complexity detection), as it results in vanishingly small inter-symbol and multi-user interference at each UT.
The complete block diagram of
the downlink transmitter at the BS is shown in Fig.~\ref{tx_fig} (the operations in (\ref{eqn14}) are implemented in the
shaded block).  

We note that the proposed precoder in (\ref{eqn14}) is different from the OTFS massive MIMO
precoder proposed in \cite{Zemen2019}. This is because, in (\ref{eqn14}) we perform spatial precoding in the DD domain whereas
in \cite{Zemen2019} beamforming is performed only in the TF domain. As we shall see later in Section \ref{seclcd},
the proposed precoder allows for separate DD domain detection of each information symbol which has a complexity of
only $O(MN \log(MN))$ when compared to the detection complexity of at least $O(M^2 N^2)$ in \cite{Zemen2019}.
The average transmitted energy is given by

{\vspace{-4mm}
\small
\begin{eqnarray}
\label{eqn16}
{\mathbb E}_{{\bf u}_s, {\bf g}_{s} }\left[ \sum\limits_{q=1}^Q \int\limits_{-\tau_{max}}^{NT}   \hspace{-3mm} \vert  x_q(t) \vert^2 \, dt  \right] & \hspace{-3mm}  = &  \hspace{-3mm}   {\mathbb E}_{{\bf u}_s, {\bf g}_{s} } \hspace{-1mm} \left[ \sum\limits_{q=1}^Q \int\limits_{0}^{NT}   \hspace{-2mm} \vert  x_q(t) \vert^2 \, dt  \right] \nonumber \\
& \hspace{-15mm}  & \hspace{-12mm}  +  \, {\mathbb E}_{{\bf u}_s, {\bf g}_{s} } \hspace{-1mm} \left[ \sum\limits_{q=1}^Q \int\limits_{-\tau_{max}}^{0}   \hspace{-2mm} \vert  x_q(t) \vert^2 \, dt  \right]
\end{eqnarray}
\normalsize}
where the expectation is w.r.t. ${\bf u}_s, {\bf g}_{s}, s=1,2,\cdots, K$ (see (\ref{gsvecdef})).
From (\ref{eqn5}), (\ref{gtxpulse_eqn}) and (\ref{eqn6}) it follows that 

{\vspace{-4mm}
\small
\begin{eqnarray}
\label{tenergyeqn}
{\mathbb E}_{{\bf u}_s, {\bf g}_{s} } \hspace{-1.5mm} \left[ \sum\limits_{q=1}^Q \hspace{-1mm} \int\limits_{0}^{NT}  \hspace{-2mm} \vert  x_q(t) \vert^2 \, dt  \right] & \hspace{-3mm}  = & \hspace{-3mm} {\mathbb E}_{{\bf u}_s, {\bf g}_{s} } \hspace{-1.5mm} \left[  \sum\limits_{q=1}^Q \sum\limits_{n=0}^{N-1}\sum\limits_{m=0}^{M-1} \vert X_q[n,m] \vert^2 \right] \nonumber \\
&  \hspace{-3mm} = &   \hspace{-3mm} {\mathbb E}_{{\bf u}_s, {\bf g}_{s} } \left[   \sum\limits_{q=1}^Q \sum\limits_{k=0}^{N-1}\sum\limits_{l=0}^{M-1} \vert x_q[k,l] \vert^2 \right] \nonumber \\
& = & \sum\limits_{q=1}^{Q}  {\mathbb E}_{{\bf u}_s, {\bf g}_{s} } \left[ \Vert  {\bf x}_q \Vert^2 \right]
\end{eqnarray}
\normalsize}
where the precoded DD domain symbol vector ${\bf x}_q$ is given by (\ref{eqn14}).
From (\ref{eqn14}) it is clear that

{\vspace{-4mm}
\small
\begin{eqnarray}
\label{eqn15}
 \sum\limits_{q=1}^{Q}  {\mathbb E}_{{\bf u}_s, {\bf g}_{s} } \hspace{-1.5mm} \left[ \Vert  {\bf x}_q \Vert^2 \right]   & \hspace{-3mm}  =  \hspace{-3mm} & \sum\limits_{q=1}^{Q} \operatorname{Tr}\Bigg(\mathbb{E}_{{\bf u}_s, {\bf g}_{s} } \Big[\mathbf{x}_q\,\mathbf{x}_q^H\Big] \Bigg) \nonumber \\
& \mya & \frac{E_T}{\eta} \, \sum\limits_{q=1}^Q \sum\limits_{s=1}^K \, \operatorname{Tr}\Bigg({\mathbb E}_{{\bf g}_{s} } \left[{\bf H}^H_{q,s}\,{\bf H}_{q,s} \right] \Bigg) \nonumber \\
& \hspace{-50mm}  \myb &  \hspace{-28mm} \frac{E_T}{\eta}  \sum\limits_{q=1}^Q \sum\limits_{s=1}^K \sum\limits_{i_1=1}^{L_s}  \sum\limits_{i_2=1}^{L_s} {\mathbb E}_{{\bf g}_{s} } \left[  h_{q,s,i_1}^* h_{q,s,i_2} \right] \operatorname{Tr}\Bigg( {\bf A}_{s,i_1}^H {\bf A}_{s,i_2}\Bigg) \nonumber \\
& \myc &  \frac{M N E_T}{\eta}  \sum\limits_{q=1}^Q \sum\limits_{s=1}^K \sum\limits_{i=1}^{L_s} {\mathbb E}_{{\bf g}_{s} } \left[  \vert g_{s,i} \vert^2 \right]  \nonumber \\
& = & E_T \frac{Q M N \sum\limits_{s=1}^K \sum\limits_{i=1}^{L_s} \beta_{s,i}}{\eta} \, = \, E_T
\end{eqnarray}
\normalsize}
where step (a) follows from the fact that $u_s[k,l] \, \sim \, \mbox{\small{i.i.d.}} \, {\mathcal C} {\mathcal N}(0,1)$.
Step (b) follows from the R.H.S. of (\ref{eq_channel}). 
Step (c) follows from the expression for $h_{q,s,i}$ in (\ref{eqn3}) and the fact that $g_{s,i}, s=1,2,\cdots, K, i=1,2,\cdots,L_s$ are i.i.d.
We have also used the fact that ${\bf A}_{s,i}^H {\bf A}_{s,i} = {\bf I}$ (see (\ref{AsiHAsi})). The last step follows from the definition of $\eta$ in (\ref{eqn14}).
Similarly, from (\ref{eqn3}), (\ref{eqn5}), (\ref{gtxpulse_eqn}), (\ref{eqn6}), (\ref{eq_channel}) and (\ref{eqn14}) it can be shown that

{\vspace{-5mm}
\small
\begin{eqnarray}
\label{TEnergyeqn1}
{\mathbb E}_{{\bf u}_s, {\bf g}_{s} } \hspace{-1mm} \left[ \sum\limits_{q=1}^Q \int\limits_{-\tau_{max}}^{0}   \hspace{-2mm} \vert  x_q(t) \vert^2 \, dt  \right] & = & E_T  \frac{\tau_{max}}{NT}
\end{eqnarray}
\normalsize}
and therefore using (\ref{tenergyeqn}), (\ref{eqn15}) and (\ref{TEnergyeqn1}) in (\ref{eqn16}), we get

{\vspace{-5mm}
\small
\begin{eqnarray}
\label{TEnergyeqn2}
{\mathbb E}_{{\bf u}_s, {\bf g}_{s} } \hspace{-1mm} \left[ \sum\limits_{q=1}^Q \int\limits_{-\tau_{max}}^{NT}   \hspace{-2mm} \vert  x_q(t) \vert^2 \, dt  \right] & \hspace{-2mm}  = &  \hspace{-2mm} E_T \left( 1 +  \frac{\tau_{max}}{NT} \right).
\end{eqnarray}
\normalsize}
The proposed precoder in (\ref{eqn14}) involves matrix multiplications (${\bf H}_{q,s}^H {\bf u}_s \,,\, s=1,2,\cdots, K$) for generating the DD domain transmit signal at each BS antenna. In Appendix \ref{appendixRavi}, from the second equation in (\ref{asikleqn}) it is clear that each column of
${\bf A}_{s,i}$ has only $N$ non-zero elements out of $MN$ elements. From (\ref{eq_channel}) it then follows that each column of ${\bf H}_{q,s}$ has at most $NL_s$ non-zero elements. Hence the complexity of the proposed precoder is $O\left(QMN^2\sum\limits_{s=1}^K L_s\right)$ for each OTFS frame. Note that
the precoding complexity increases only {\em linearly} with the number of BS antennas $Q$ and the number of UTs $K$.  

Using (\ref{eqn14}) in (\ref{eqn10}), the received vector of DD domain symbols at the $s$-th UT is

{\vspace{-4mm}
\small
\begin{eqnarray}
\label{eqn18}
{\widehat {\bf x}_s} & = & \sum\limits_{q=1}^Q {\bf H}_{q,s} \left(  \sqrt{\frac{E_T}{\eta}}  \sum\limits_{s'=1}^K {\bf H}_{q,s'}^H {\bf u}_{s'} \right)  \, + \, {\bf w}_s \nonumber \\
& = & \sqrt{\frac{E_T}{\eta}} {\bf G}_{s,s}  {\bf u}_s  \, + \, \sqrt{\frac{E_T}{\eta}}  \sum\limits_{s' = 1 \atop s' \ne s}^K  {\bf G}_{s,s'}  {\bf u}_{s'} \, + \,  {\bf w}_s \,,\, \nonumber \\
{\bf G}_{s,s'}  & \Define  & \sum\limits_{q=1}^Q {\bf H}_{q,s} {\bf H}_{q,s'}^H.
\end{eqnarray} 
\normalsize}
Using (\ref{eq_channel}) in (\ref{eqn18}) we get

{\vspace{-4mm}
\small
\begin{eqnarray}
\label{eqn19}
\hspace{-2mm} \mathbf{G}_{s,s'} & \hspace{-3mm} =&  \hspace{-3mm} \sum_{q = 1}^{Q}\mathbf{H}_{q,s} \,\mathbf{H}^H_{q,s'} 
%  &  \hspace{-3mm} =&  \hspace{-3mm} \sum_{q=1}^{Q} \Big(\sum_{i=1}^{L_s} h_{q,s,i} \, \mathbf{A}_{s,i}\Big) \Big(\sum_{k=1}^{L_{s'}}h_{q,s^{\prime},k} \, \mathbf{A}_{s^{\prime},k}\Big)^H\nonumber\\
   =  \sum_{q=1}^{Q}\sum_{i=1}^{L_s}\sum_{k=1}^{L_{s'}}h_{q,s,i}h^*_{q,s',k} \mathbf{A}_{s,i} \mathbf{A}^H_{s',k}.
\end{eqnarray}
\normalsize}
In (\ref{eqn18}), the relation between the transmitted information symbol vector $ {\bf u}_s$ and the received DD domain signal vector ${\widehat {\bf x}_s}$
is similar to that of a $MN \times MN$ MIMO system, for which the
largest information rate (i.e., $I({\widehat {\bf x}_s} ;  {\bf u}_s)$) is achieved by the
Minimum Mean Squared Error Estimation - Successive Interference Cancellation (MMSE-SIC) detector \cite{DTse}.
Here $I(x ; y)$ denotes the mutual information between the random variables $x$ and $y$ \cite{Cover}.     
Therefore, for given channel parameters ${\mathcal P} = \left\{ \{L_s\}_{s=1}^K, \{\theta_{s,i}, \phi_{s,i}, \tau_{s,i}, \nu_{s,i}, g_{s,i} \}_{s=1, i=1}^{K, L_s} \right\}$,
the SE achieved by the proposed precoder in (\ref{eqn14}) with
the optimal MMSE-SIC detector is given by (\ref{Cvdefeqn}) (see top of next page).
In (\ref{Cvdefeqn}), the factor $MN\left(1 + \frac{\tau_{max}}{NT} \right)$ is the total time-bandwidth product i.e., $M \Delta f \left(\tau_{max} + NT  \right)$. 
\begin{figure*}
{\vspace{-9mm}
\small
\begin{eqnarray}
\label{Cvdefeqn}
C_s({\mathcal P}) & \Define &  \frac{I({\widehat {\bf x}_s} ;  {\bf u}_s)}{MN \left( 1 + \frac{\tau_{max}}{NT} \right)} \, = \, \frac{1}{MN \left( 1 + \frac{\tau_{max}}{NT} \right)} \,   \log_2 {\Bigg \vert}  {\bf I} \, + \, \frac{E_T}{\eta N_o} {\bf G}_{s,s} {\bf G}_{s,s}^H   {\Bigg (} {\bf I} + \frac{E_T}{\eta N_o}  \sum\limits_{s' = 1, \atop s' \ne s}^K {\bf G}_{s,s'} {\bf G}_{s,s'}^H  {\Bigg )}^{-1}  {\Bigg \vert}.
\end{eqnarray}
\begin{eqnarray*}
\hline
\end{eqnarray*}
\vspace{-8mm}
\normalsize}
\end{figure*}
The sum SE achieved by the proposed precoder is therefore given by

{\vspace{-4mm}
\small
\begin{eqnarray}
\label{speceffeqn}
C({\mathcal P}) & \Define & \sum\limits_{s=1}^K  C_s({\mathcal P}).
\end{eqnarray}
\normalsize}
From (\ref{TEnergyeqn2}) we know that the total transmitted energy is $E_T(\tau_{max} + NT)/(NT)$ and therefore the average total transmitted power is $E_T/NT$ as the total
duration of the OTFS frame is $(\tau_{max} + NT)$. At the receiver, the AWGN power is $M \Delta f N_o$ since $N_o$ is the PSD of the AWGN and $M \Delta f$ is the
total communication bandwidth. Subsequently, in this paper we denote the ratio of the average total transmitted power to the AWGN power at the receiver by

{\vspace{-4mm}
\small
\begin{eqnarray}
\label{rhodefeqn}
\rho & \Define & \frac{E_T/(NT)}{M \Delta f N_o} \, = \, \frac{E_T}{M N N_o}
\end{eqnarray}
\normalsize}
which follows from the fact that $T \Delta f = 1$.

The complexity of the optimal MMSE-SIC detector is however $O(M^4 N^4)$ and therefore, achieving the SE in (\ref{Cvdefeqn}) is challenging when $M$ and/or $N$ is large. 
In the next section, we therefore propose a low complexity detector which is shown to achieve a sum SE close to that of the optimal MMSE-SIC detector when the number of BS antennas ($Q$) is large.

\section{Low Complexity Detection of Received Multi-user OTFS Signal}
\label{seclcd}
In this section, we propose an alternate low complexity detector (LCD) where each information symbol in the DD domain is detected separately.   
In the following we propose the LCD for $u_{s,r}$ (the $r$-th element of ${\bf u}_s$). Let ${\widehat x}_{s,r}$ denote the $r$-th element of
the DD domain vector ${\widehat {\bf x}_s}$ $(r=1,2,\cdots, MN \,,\, s=1,2,\cdots,K)$ received at the $s$-th UT. From (\ref{eqn18}) it follows that

\begin{figure}[t]
\vspace{-0.4 cm}
\centering
\includegraphics[width= 3.1 in, height= 2.0 in]{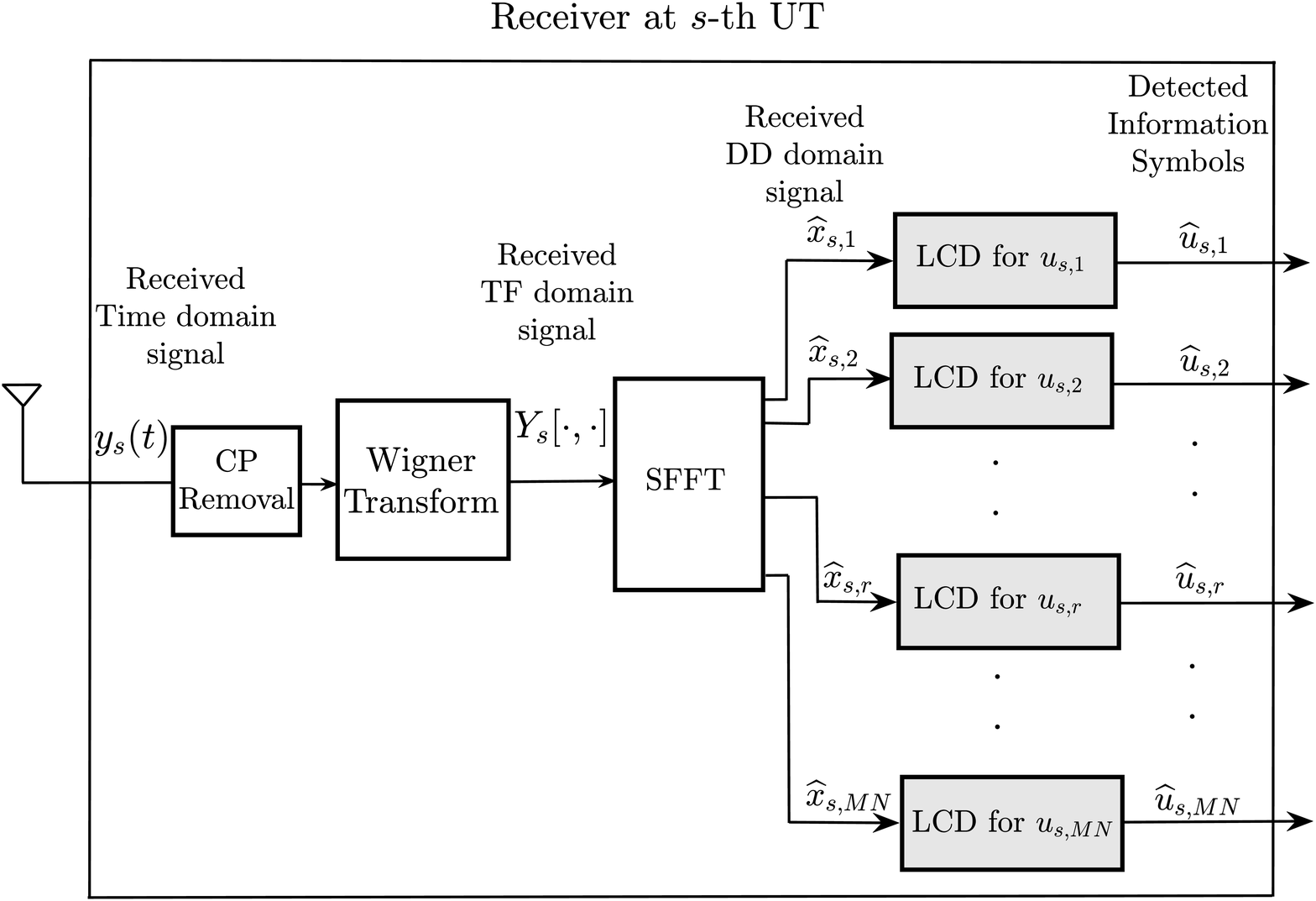}
\vspace{-0.1 cm}
\caption{Block diagram of the receiver at the $s$-th UT with the proposed LCD detector.} 
\vspace{-0.1cm}
\label{rx_fig}
\end{figure}

{\vspace{-4mm}
\small
\begin{eqnarray}
\label{eqn20}
{\widehat x}_{s,r} & = & \sqrt{\frac{E_T}{\eta}} \, \gamma_{s,s,r,r} \, u_{s,r} \, + \,   {\Tilde w}_{s,r} \nonumber  \,,\, \\
 {\Tilde w}_{s,r} &  \hspace{-3mm}  \Define &    \underbrace{\sqrt{\frac{E_T}{\eta}}\sum_{p=1  \atop p \neq r}^{MN} \gamma_{s,s,r,p}\,u_{s,p} }_{\mbox{\tiny{Inter-symbol Interference}}} \nonumber \\
& &  \, + \,  \underbrace{\sqrt{\frac{E_T}{\eta}}\sum\limits_{s' = 1 \atop s' \ne s}^K \sum\limits_{p=1}^{MN}  \gamma_{s,s',r,p}\,u_{s',p}}_{\mbox{\tiny{Multi-user Interference}}}   +  \underbrace{{w_{s,r}}}_{\mbox{\tiny{AWGN}}} 
\end{eqnarray}
\normalsize}
where $\gamma_{s,s',r,p}$ denotes the element of ${\bf G}_{s,s'}$ in its $r$-th row and $p$-th column. Here $w_{s,r}$ denotes the $r$-th element of ${\bf w}_s$.
In (\ref{eqn20}), ${\Tilde w}_{s,r}$ is the effective noise and interference and $\sqrt{\frac{E_T}{\eta}} \gamma_{s,s,r,r} \, u_{s,r} $ is the useful signal. The channel in (\ref{eqn20}) is effectively a single-input single-output (SISO) channel with $u_{s,r}$ as input and ${\widehat x}_{s,r}$ as output. Since $r=1,2,\cdots, MN$,  for each UT we therefore have $MN$ sub-channels, one for each DDRE location (for example, from (\ref{eqn9}) and (\ref{eqn20}) it is clear that $u_{s,r}$ is transmitted and detected on the $(k=\lfloor (r-1)/M \rfloor, l=[r-1]_{M})$-th DDRE). At the encoder, different information codebooks are used in principle for generating information codewords for each such sub-channel of each UT. If the total communication time is limited to $F$ OTFS frames and if $(c_{s,r,1}, c_{s,r,2}, \cdots, c_{s,r,f}, \cdots, c_{s,r,F})$ denotes the information codeword for the $s$-th UT to be transmitted on the $r$-th sub-channel, then the information symbol $c_{s,r,f}$ is transmitted in the $f$-th OTFS frame on the $(k=\lfloor (r-1)/M \rfloor, l=[r-1]_{M})$-th DDRE (i.e., in the $f$-th OTFS frame, $u_{s,r} = c_{s,r,f}$). In the proposed LCD detector, the UT performs \emph{separate} decoding for each SISO channel which greatly reduces the complexity when compared to MMSE-SIC. For example, if the received DD domain symbol in the $f$-th OTFS frame on the $(k=\lfloor (r-1)/M \rfloor, l=[r-1]_{M})$-th DDRE is denoted by ${\widehat x}_{s,r,f}$, then the codeword $(c_{s,r,1}, c_{s,r,2}, \cdots, c_{s,r,f}, \cdots, c_{s,r,F})$ is decoded from the received symbol vector $({\widehat x}_{s,r,1}, {\widehat x}_{s,r,2}, \cdots, {\widehat x}_{s,r,F})$. In Fig.~\ref{rx_fig} we show the receiver at the $s$-th UT with the proposed LCD detector. Separate LCD detectors are shown for each of the $MN$ sub-channels (see the shaded blocks in Fig.~\ref{rx_fig}).   

Due to separate symbol detection, the complexity of the LCD detector is $O(MN)$ as there are $MN$ information symbols in one OTFS frame. Since the complexity of
the TF to DD domain conversion in (\ref{eqn8}) is $O(MN \log(MN))$, the overall complexity of LCD along with the TF to DD domain converter is only $O(MN \log(MN))$.
Note that this complexity is independent of the number of channel paths and is smaller than the complexity of the detector in \cite{Zemen2019} (which is at least $O(M^2N^2)$).
In the following we derive an expression for the achievable sum SE of the proposed LCD detector.  
\begin{theorem}
\label{thm1}
For a given set of channel parameters ${\mathcal P} = \left\{ \{L_s\}_{s=1}^K, \{ \theta_{s,i}, \phi_{s,i}, \tau_{s,i}, \nu_{s,i}, g_{s,i} \}_{s=1, i=1}^{K, L_s} \right\}$, the SE achieved by the $s$-th UT with the proposed LCD detector is given by

{\vspace{-5mm}
\small
\begin{eqnarray}
\label{eqnlem1}
R_s({\mathcal P}) & \Define & \frac{1}{MN \left(1 + \frac{\tau_{max}}{NT}  \right)} \sum\limits_{r=1}^{MN} I_{s,r}({\mathcal P})  \,\,,\,\, \nonumber \\
I_{s,r}({\mathcal P})   & \Define &   \log_2\left(  1 +  \mbox{SINR}_{s,r} ({\mathcal P}) \, \right)  \,\,,\,\, \nonumber \\
\mbox{SINR}_{s,r} ({\mathcal P})& \hspace{-3mm}  \Define &  \hspace{-3mm} \frac{\left\vert \gamma_{s,s,r,r} \right\vert^2}{\underbrace{ \frac{\eta}{\rho M N }}_{\mbox{\tiny{AWGN}}} 
+ \underbrace{\sum\limits_{p=1  \atop p \neq r}^{MN}\,\big| \gamma_{s,s,r,p} \big|^2}_{\mbox{\tiny{Inter-symbol Interference (ISI)}}}  +  \underbrace{\sum\limits_{s' = 1 \atop s' \ne s}^K \sum\limits_{p=1}^{MN}  \left\vert  \gamma_{s,s',r,p}  \right\vert^2  }_{\mbox{\tiny{Multi-user Interference (MUI)}}}   }. \nonumber \\
\end{eqnarray}
\normalsize}
\end{theorem}
\begin{IEEEproof}
For given channel parameters ${\mathcal P}$, the channel in (\ref{eqn20}) is a SISO channel with complex Gaussian information symbol $u_{s,r}$ and statistically independent effective noise and interference ${\Tilde w}_{s,r}$.
The statistical independence is due to i.i.d. information symbols $u_{s,p} \,,\, p=1,2,\cdots, MN$ and independent AWGN. The achievable rate expression in (\ref{eqnlem1}) then follows from the capacity of such SISO channels \cite{DTse}.
%Here we have also used the fact that $N_0/E_T = 1/(\rho M N)$ (see (\ref{rhodefeqn})).
In (\ref{eqnlem1}), the factor $MN\left(1 + \frac{\tau_{max}}{NT} \right)$ is the total time-bandwidth product i.e., $M \Delta f \left(\tau_{max} + NT  \right)$.
\end{IEEEproof}
The sum SE achieved with the proposed LCD detector is therefore given by

{\vspace{-4mm}
\small
\begin{eqnarray}
\label{rsumlcdeqn}
R({\mathcal P}) & \Define & \sum\limits_{s=1}^K R_s({\mathcal P}).
\end{eqnarray}
\normalsize}
\begin{corollary}
\label{cor11}
For fixed system and channel parameters i.e., fixed $(B_c, T_c, \Delta f, d/\lambda, {\mathcal P})$, as the URA size increases i.e., $(Q_h \rightarrow \infty, Q_v \rightarrow \infty)$ with constant $\rho \, Q $,
an approximation to $I_{s,r}$ is given by (\ref{eqn269}) (see top of next page).
From this approximation, it is clear that in the large $(Q_h,Q_v)$ and small $\rho$ regime, the achievable information rate of the proposed precoder with the LCD detector is limited only by AWGN, and becomes \emph{independent} of inter-symbol and multi-user interference.
\begin{figure*}
{\vspace{-9mm}
\small
\begin{eqnarray}   
\label{eqn269}
I_{s,r} & \myapproxa &  \log_2 \left[ 1 + \left\{ {\rho Q \left( \sum\limits_{i=1}^{L_s} \vert g_{s,i} \vert^2 \right)^2} {\Bigg /} {\sum\limits_{{\Tilde s}=1}^K \sum\limits_{i=1}^{L_{\Tilde s}}  \beta_{{\Tilde s},i} }  \right\} \right]  \,\,,\,\, \mbox{\small{Large}} \,\, (Q_h, Q_v) \,\, \mbox{\small{approximation with constant}}  \,\, \rho \, Q.
\end{eqnarray}
\vspace{-3mm}
\begin{eqnarray*}
\hline
\end{eqnarray*}
\normalsize}  
\vspace{-7mm}
\end{figure*}
\end{corollary}

\begin{IEEEproof}
Step (a) in (\ref{eqn269}) follows by substituting the large $(Q_h,Q_v)$ approximation of $\gamma_{s,s',r,p}/Q$, $\gamma_{s,s,r,r}/Q$, $\gamma_{s,s,r,p}/Q$ from Appendices \ref{egvvprime}, \ref{appendixA}, \ref{appendixC} and the expression for $\eta$ from (\ref{eqn14}) into the expression for $\mbox{SINR}_{s,r}$ in (\ref{eqnlem1}). From Appendices \ref{egvvprime} and \ref{appendixC} we know that
for large $(Q_h,Q_v)$ (constant $ \rho Q$), $\gamma_{s,s',r,p}/Q \approx 0 \, (s \ne s')$ and for $p \ne r$, $\gamma_{s,s,r,p}/Q \approx 0$. Using this fact in the expression for $\mbox{SINR}_{s,r}$ in (\ref{eqnlem1}), it is clear that
in the large antenna and small $\rho$ regime, the effective noise and interference power (i.e., denominator in the R.H.S. of the expression for $\mbox{SINR}_{s,r}$ in (\ref{eqnlem1})) is dominated by AWGN.       
\end{IEEEproof}

The following result shows that in the large $(Q_h,Q_v)$ and small $\rho$ regime, the proposed LCD detector is near-optimal.
\begin{theorem}
\label{thm34}
For fixed system and channel parameters i.e., fixed $(B_c, T_c, \Delta f, d/\lambda, {\mathcal P})$, as the URA size increases i.e., $(Q_h \rightarrow \infty, Q_v \rightarrow \infty)$ with constant $\rho \, Q $,
the proposed LCD detector is near-optimal, i.e.,
\begin{eqnarray}
R({\mathcal P}) \approx C({\mathcal P}) \,\,,\,\, \mbox{\small{Large}} \,\, (Q_h, Q_v) \,\, \mbox{\small{Approximation}}.
\end{eqnarray}
\end{theorem}

\begin{IEEEproof}
Refer to Appendix \ref{appendixD}.
\end{IEEEproof}

\textbf{Discussion}: From Corollary \ref{cor11} and Theorem \ref{thm34} the following conclusions can be made.
\begin{enumerate}
\item From Corollary \ref{cor11} it is clear that with a large antenna array the achievable information rate with the proposed LCD detector depends on $\rho$ only through the product $\rho Q$, and therefore
in the large $(Q_h,Q_v)$ regime, for a desired achievable sum SE, the required $\rho$ will {\em decrease} as $1/Q$ with increasing $(Q_h,Q_v)$. This implies that the energy efficiency improves with increase in $Q$. This effect is the same as that observed in TF domain based OFDM massive MIMO systems with perfect channel state information (CSI) \cite{MassiveMIMOBook}.
Also, just as in TF domain based massive MIMO systems, from Corollary \ref{cor11} we also observe that even in OTFS based massive MIMO systems the effect of multi-user and inter-symbol interference vanishes with increasing number of BS antennas, i.e., in the large antenna and small $\rho$ regime the sum SE is limited only by AWGN. From the proof of Corollary \ref{cor11} it is clear that this happens due to the proposed precoder
in (\ref{eqn14}), which results in
combining of the channel matrices between the $s$-th UT and the BS (i.e., ${\bf H}_{q,s}, q=1,2,\cdots, Q$) into the effective DD domain channel matrix ${\bf G}_{s,s}   =   \sum\limits_{q=1}^Q {\bf H}_{q,s} {\bf H}_{q,s}^H$ (see (\ref{eqn18}) and (\ref{eqn20})), whose off-diagonal entries are vanishingly smaller than its diagonal entries when $Q$ is large (i.e., vanishing inter-symbol interference with increasing $Q$). Further, due to the proposed precoder, the channel matrices between the BS and the other UTs (i.e., $\mathbf{H}_{q,s'}, q =1,2,\cdots, Q, s' \ne s$) combine with the channel matrices for the $s$-th UT resulting in the matrices ${\bf G}_{s,s'}   =  \sum_{q = 1}^{Q}\mathbf{H}_{q,s} \,\mathbf{H}^H_{q,s'} $ which cause multi-user interference (see (\ref{eqn18}) and (\ref{eqn20})), but whose entries are vanishingly small (compared to the diagonal entries of ${\bf G}_{s,s}$) with increasing $Q$. From Theorem \ref{thm34} it is further clear that in the large antenna and small $\rho$ regime, the proposed LCD detector is near-optimal i.e., $R({\mathcal P}) \approx C({\mathcal P})$.       
\end{enumerate}

\section{Numerical Simulations}
For numerical simulations, we consider a single circular cell of radius $5$ Km with a $Q= Q_h Q_v$ antenna BS at its centre serving
$K$ single antenna UTs in the downlink. The UTs are uniformly distributed in the cell except in a circular region of radius $35$ m around the BS.
We consider the Rural Macro Non Line-of-Sight (RMa-NLOS) scenario which is expected to be deployed for wide area coverage (e.g., supporting high speed vehicles) \cite{tr3gpp}.
The path-loss model\footnote{\footnotesize{As a reference, we consider the path-loss to a cell-edge UT to be unity and therefore the variable $\rho$ denotes the average received signal-to-noise ratio (in the time-domain) at a cell-edge UT.}},  
the power delay profile (i.e., number of paths, path delays and their relative power gains) and the angles of departure are
described in Section $7.4.1$ and Table $7.5-6$ (Part-$2$) in \cite{tr3gpp}.\footnote{\footnotesize{As the cell-size ($5$ Km radius) is not small, we consider a mean delay spread of $0.37 \mu s$.}} 
The Doppler shift for each path is modelled as $\nu_{s,i} = \nu_{max} \cos(\alpha_{s,i})$ where $\alpha_{s,i}$ are distributed uniformly in $[0 \,,\, 2 \pi)$. Further, $\alpha_{s,i}$ are statistically independent for different $s=1,2,\cdots,K$ and $i=1,2,\cdots, L_s$.
The spacing between adjacent antenna elements in the URA is taken to be half of the carrier wavelength. The maximum possible channel path delay is taken to be $\tau_{max} = 4.7 \mu s$, which is the CP size in 4G-LTE systems. 
The system parameters $(B_c, \Delta f, T_c, M, N, f_c)$ are tabulated in Table-\ref{tab1sysparam}.           

\begin{table}[h!]
\begin{center}
\caption{System Parameters}
\label{tab1sysparam}
\begin{tabular}{|c|c|}
\hline
\mbox{Parameter} & \mbox{Value} \\
\hline
\mbox{Carrier Frequency}\, $(f_c)$ & $4.8$ GHz \\
\hline
\mbox{Sub-carrier Spacing}\, $(\Delta f)$ & $15$ KHz \\
\hline
$M$, $N$ & $M = 330$, $N = 4$ \\
\hline
\mbox{Bandwidth} $(B_c)$ & $B_c = M \Delta f = 4.95$ MHz \\
\hline
\mbox{OTFS Frame Duration} $(T_c)$ & $T_c = N/\Delta f = 0.266$ ms \\
%\hline
%\mbox{BS and UT height} $h_{BS}, h_{UT}$ & $h_{BS} =  35$ m, $h_{UT} = 1.5$ m \\
\hline
\end{tabular}
\end{center}
\end{table}

\begin{figure}[t]
\vspace{-0.7 cm}
\hspace{-0.2 in}
\centering
\includegraphics[width= 3.0 in, height= 2.1 in]{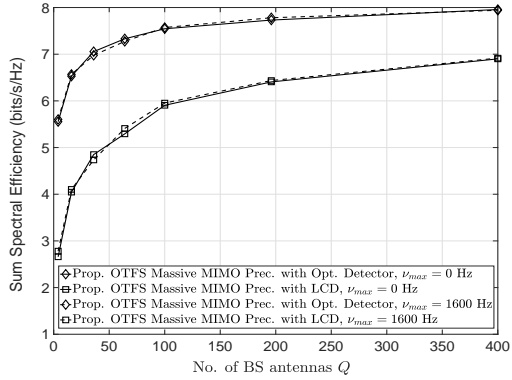}
\vspace{-0.08 cm}
\caption{Sum SE (bits/s/Hz) versus number of BS antennas $Q$ (constant $ \rho \, Q = 0.1$).} 
\vspace{-0.1cm}
\label{nearopt_fig}
\end{figure}
In Fig.~\ref{nearopt_fig}, we plot the sum SE achieved by the proposed OTFS based precoder with two different detectors, i.e., i) optimal MMSE-SIC detector, and ii) proposed LCD detector.
The SEs are plotted as a function of increasing URA size $Q$ (with $Q_h = Q_v$) and a constant $\rho Q = 0.1$. The plotted SEs are averaged over the statistics of the  
channel parameters ${\mathcal P} = \left\{ \{L_s\}_{s=1}^K, \{\beta_{s,i}, \theta_{s,i}, \phi_{s,i}, \tau_{s,i}, \nu_{s,i} \}_{s=1, i=1}^{K, L_s} \right\}$. Therefore, the plotted SE is ${\mathbb E}[C({\mathcal P})]$ and ${\mathbb E}[R({\mathcal P})]$ respectively for the optimal detector and the proposed LCD detector (see (\ref{speceffeqn}) and (\ref{rsumlcdeqn})). There are $K=4$ UTs and we consider two scenarios, i) where the maximum Doppler shift for each UT is $\nu_{max} = 1600$ Hz which corresponds to a mobile speed of $360$ Km/hr at a carrier frequency of $f_c = 4.8$ GHz, and ii) where the UTs are stationary (i.e., $\nu_{max} = 0$). The other system parameters are as in Table-\ref{tab1sysparam}. From Fig.~\ref{nearopt_fig} it is clear that the gap between the SE achieved by the LCD detector and that achieved by the optimal detector {\em diminishes} with increasing number of BS antennas. This supports our analysis in Theorem \ref{thm34}. It is also observed that for both the detectors,
the SE at a high mobile speed of $360$ Km/hr (i.e., $\nu_{max} = 1600$ Hz) is almost same as the SE at zero mobile speed, i.e., the SE is almost {\em invariant} to channel induced Doppler shift.

\begin{figure}[t]
\vspace{-0.7 cm}
\hspace{-0.2 in}
\centering
\includegraphics[width= 3.0 in, height= 2.1 in]{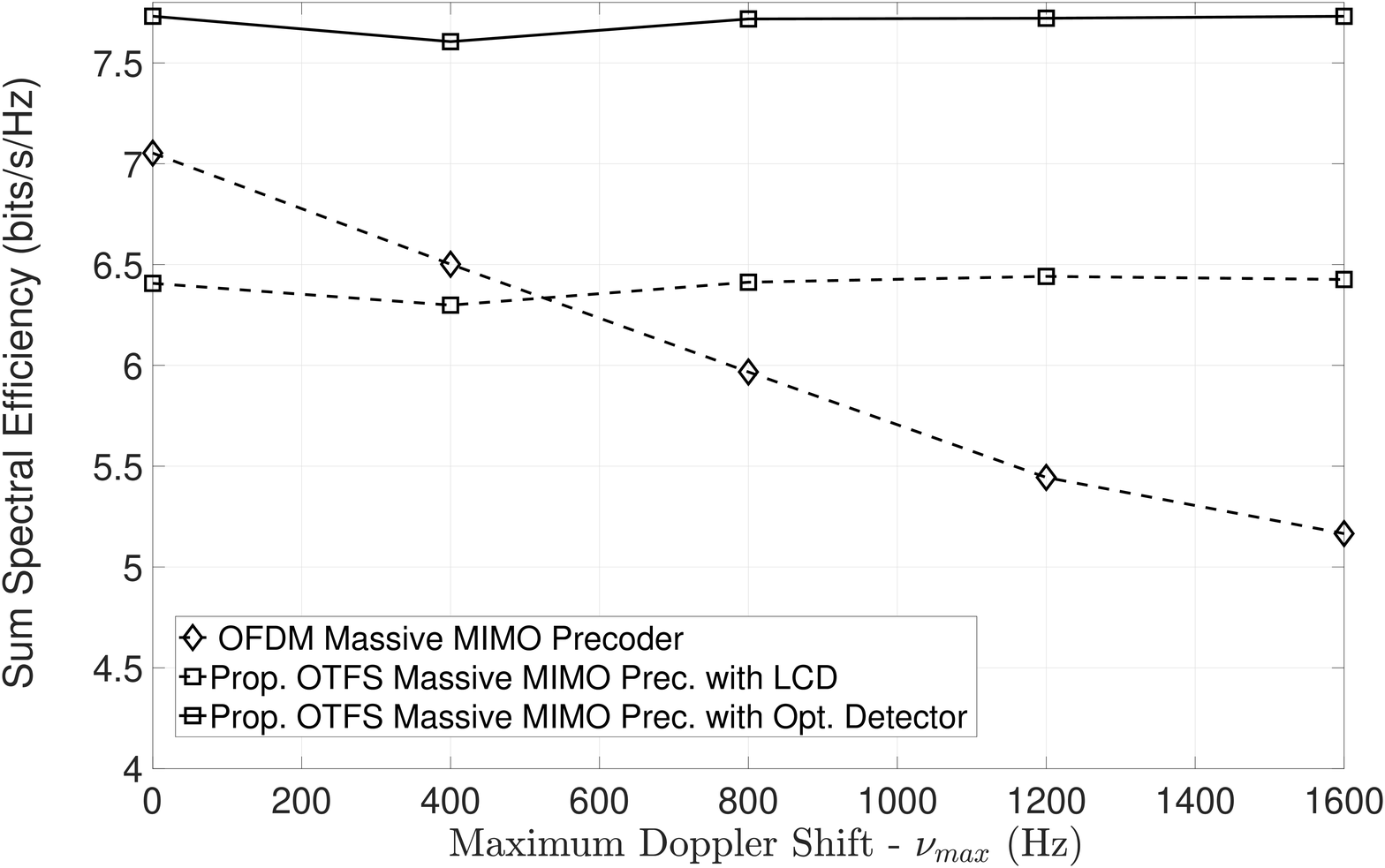}
\vspace{-0.1 cm}
\caption{Sum SE (bits/s/Hz) versus maximum Doppler shift $\nu_{max}$ (Hz).} 
\vspace{-0.08cm}
\label{lcd_fig}
\end{figure}

In Fig.~\ref{lcd_fig}, we plot the average sum SE achieved by the proposed downlink precoder with the proposed LCD detector
(i.e., ${\mathbb E}[R({\mathcal P})]$), for a OTFS multi-user massive MIMO downlink system as a function of increasing
maximum Doppler shift $\nu_{max}$.
The BS has a URA of size $Q_h = Q_v = 14$, i.e., $Q = 196$ antennas, $\rho Q = 0.1$ and $K=4$ UTs. The other system
parameters are as in Table-\ref{tab1sysparam}.     
In Fig.~\ref{lcd_fig}, we also plot the SE achieved by the proposed downlink precoder with the optimal detector in the DD domain (i.e., ${\mathbb E}[C({\mathcal P})]$), and the SE achieved
by an OFDM multi-user massive MIMO downlink system (with Maximum Ratio Transmission precoding in the TF domain \cite{MassiveMIMOBook}). The sub-carrier spacing for the OFDM massive MIMO system is $\Delta f = 15$ KHz. From the figure it is clear that the SE achieved by OFDM massive MIMO degrades significantly at high mobile speed whereas the SE achieved by the proposed precoder (with either the LCD detector or the optimal detector) does not degrade. At high mobile speed, the proposed DD domain based downlink precoder with the LCD detector achieves significantly {\em better} SE when compared to that achieved by an OFDM massive MIMO system.             

In Table-\ref{tab21} we list the sum SE achieved by the proposed OTFS based downlink precoder with the LCD detector and that
achieved by the OFDM massive MIMO system as a function of varying $\rho$ when the BS has $Q = 196$ BS antennas ($Q_h = Q_v = 14$) and there are $K=4$ UTs.
From the table, it is observed that for all considered values of $\rho$, at high Doppler spread the sum SE achieved by OTFS massive MIMO is significantly greater than that achieved by OFDM massive MIMO. Also, for all considered values of $\rho$, the sum SE achieved by the OTFS massive MIMO system
is almost invariant of the Doppler spread. 

\begin{table*}
\vspace{-6mm}
\caption{Sum SE (bits/s/Hz) vs. $(\nu_{max}, \rho Q)$, $Q_h = Q_v = 14$, $Q = Q_h Q_v = 196$, $K=4$ UTs.}
{\small
\begin{center}
\vspace{-3mm}
\begin{tabular}{||c|c|c||c|c||c|c||c|c||c|c||}
\hline
& \multicolumn{2}{|c|}{$\rho Q = -19$ dB} & \multicolumn{2}{c|}{$\rho Q = -16$ dB}  &  \multicolumn{2}{c|}{$\rho Q = -13$ dB}  &   \multicolumn{2}{c|}{$\rho Q = -10$ dB}  &  \multicolumn{2}{c|}{$\rho Q = -7$ dB}\nonumber \\
\hline
& OTFS & OFDM & OTFS & OFDM  & OTFS & OFDM & OTFS & OFDM   & OTFS & OFDM  \nonumber \\
\hline
$\nu_{max} = 0$ Hz &       $4.4$ &  $4.6$ & $5.0$ & $5.4$  & $5.7$ & $6.2$  &  $6.4$  &  $7.1$   &    $7.1$   &  $8.1$ \nonumber \\
\hline
$\nu_{max} = 400$ Hz &   $4.4$ &  $4.2$ & $5.0$ & $4.9$  & $5.6$ & $5.6$  &  $6.3$  &  $6.4$   &    $7.0$   &  $7.2$ \nonumber \\
\hline
$\nu_{max} = 800$ Hz &   $4.4$ &  $3.9$ & $5.0$ & $4.5$  & $5.7$ & $5.2$  &  $6.4$  &  $5.9$   &    $7.1$   &  $6.6$ \nonumber \\
\hline
$\nu_{max} = 1200$ Hz &   $4.5$ &  $3.7$ & $5.0$ & $4.2$  & $5.7$ & $4.8$  &  $6.4$  &  $5.5$   &    $7.1$   &  $6.2$ \nonumber \\
\hline
$\nu_{max} = 1600$ Hz &   $4.5$ &  $3.5$ & $5.0$ & $4.0$  & $5.7$ & $4.6$  &  $6.4$  &  $5.2$   &    $7.1$   &  $5.9$ \nonumber \\
\hline
\end{tabular}
\vspace{-4mm}
\end{center}
\normalsize}
\label{tab21}
\end{table*}

\begin{figure}[t]
\vspace{-0.4 cm}
\hspace{-0.2 in}
\centering
\includegraphics[width= 3.0 in, height= 2.1 in]{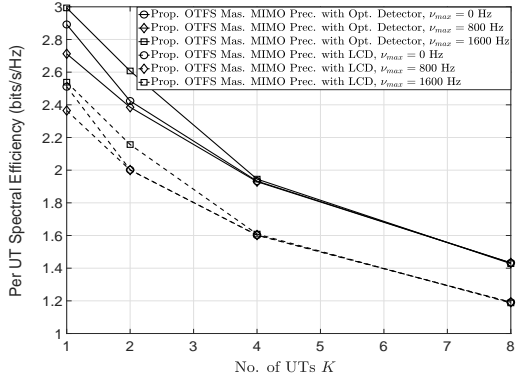}
\vspace{-0.2 cm}
\caption{Per UT SE (bits/s/Hz) versus number of UTs $K$.} 
\vspace{-0.2cm}
\label{numut_fig}
\end{figure}

In Fig.~\ref{numut_fig}, we plot the per UT SE achieved by the proposed downlink precoder with both the LCD and the optimal detector (i.e., ${\mathbb E}[R({\mathcal P})]/K$ and ${\mathbb E}[C({\mathcal P})]/K$ respectively) as a function of increasing number of UTs $K$. The BS has a URA of size $Q_h = Q_v = 14$, i.e., $Q = 196$ antennas and $\rho \, Q = 0.1$. 
%It is observed that the gap between the per UT SE achieved by the proposed LCD detector and that by the optimal detector, {\em decreases} with increase in the number of UTs.
From the figure we observe that the gap between the per UT SE achieved by the proposed LCD detector and that by the optimal detector
is around $0.4$ bits/s/Hz with $K=2$ UTs, but decreases to $0.2$ bits/s/Hz when the number of UTs is increased to $K=8$.    

\begin{figure}[t]
\vspace{-0.3 cm}
\hspace{-0.2 in}
\centering
\includegraphics[width= 3.0 in, height= 2.1 in]{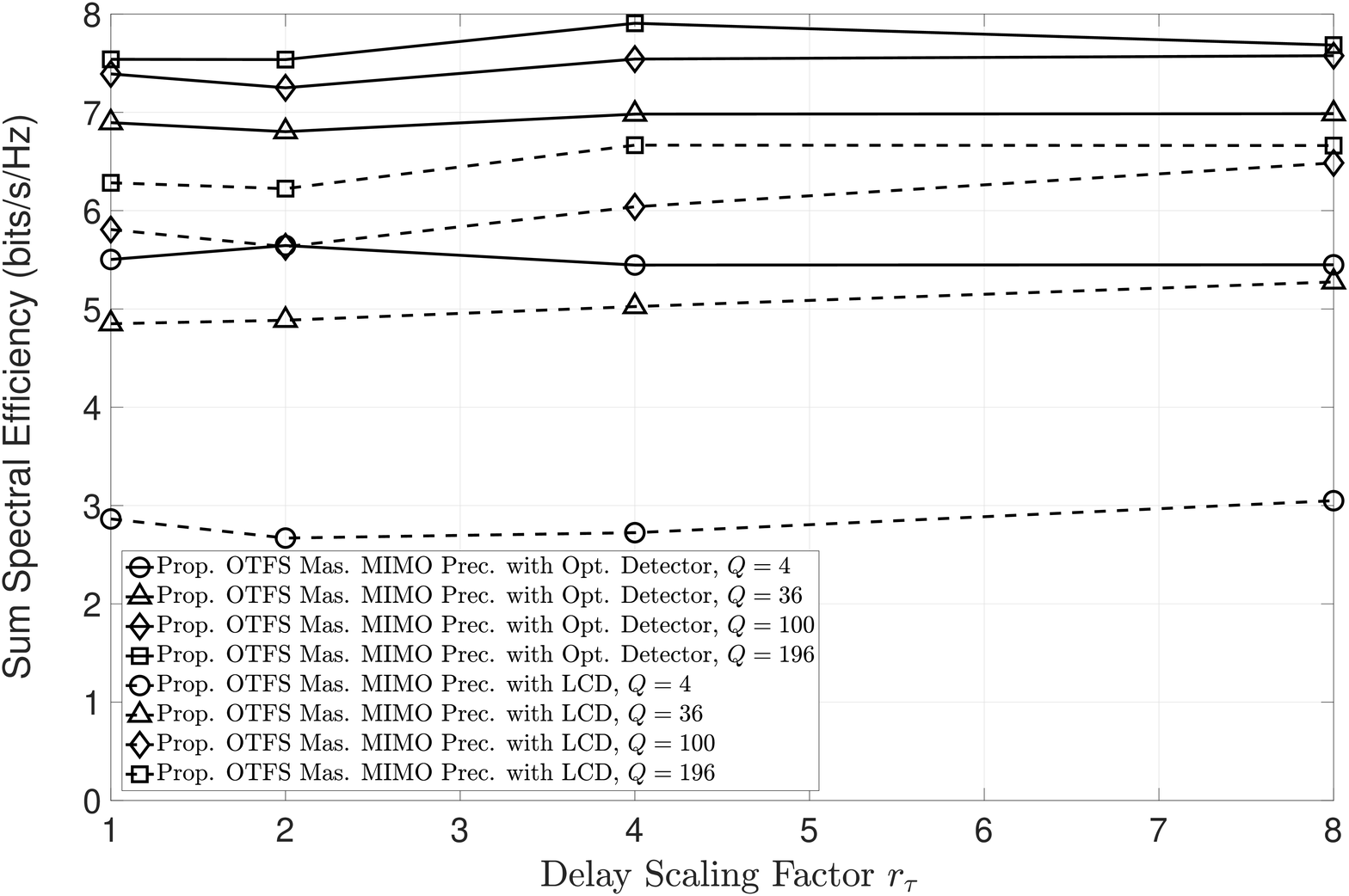}
\vspace{-0.2 cm}
\caption{Sum SE (bits/s/Hz) vs. delay scaling factor $r_{\tau}$.} 
\vspace{-0.2cm}
\label{pdppr_fig}
\end{figure}

In Fig.~\ref{pdppr_fig}, we plot the variation in the sum SE achieved by the proposed precoder as a function of varying power delay profile, but fixed
number of UTs $K=4$, fixed $\rho Q = 0.1$ and fixed $\nu_{max} = 1600$ Hz.
In the RMa-NLOS channel model \cite{tr3gpp}, the random channel path delays are modelled as
$\tau = - r_{\tau} \mu_{\tau} \log (X)$ where $X$ is uniformly distributed in the interval $[0 \,,\, 1]$.
The power gain of a path having delay $\tau$ is proportional to $e^{- \tau \frac{(r_{\tau} - 1)}{r_{\tau} \mu_{\tau}}}$.
Here $\mu_{\tau}$ is the mean delay spread and $r_{\tau}$ is the delay scaling factor which can be used to vary the power profile of the channel paths.
A value of $r_{\tau} = 1$ models a flat power profile where all channel paths have the same power irrespective of their delay.
With increasing $r_{\tau}$, the power profile becomes steeper, i.e., the power of channel paths decreases more rapidly with increasing path delay.
From Fig.~\ref{pdppr_fig} it is clear that the SE achieved by the proposed OTFS massive MIMO precoder is almost invariant of the power profile and does not degrade
when there are a large number of equal power paths (i.e., when $r_{\tau}$ is close to $1$).         

\subsection{Coded Error Rate Performance Comparison between OTFS and OFDM Massive MIMO Systems with Imperfect CSI}
\label{errorrateperf}
In this section, we present the comparison between the error
rate performance of the proposed OTFS precoder (with the LCD detector) and the
OFDM massive MIMO precoder, for the imperfect channel state information (CSI) scenario.
In both the proposed OTFS system and the OFDM based system, we perform uplink channel estimation and the estimated uplink channel is used
for downlink precoding. This strategy is used since we consider a TDD system where the uplink and downlink channels are reciprocal.

\begin{figure}[t]
\vspace{-0.2 cm}
\hspace{-0.2 in}
\centering
\includegraphics[width= 2.9 in, height= 2.0 in]{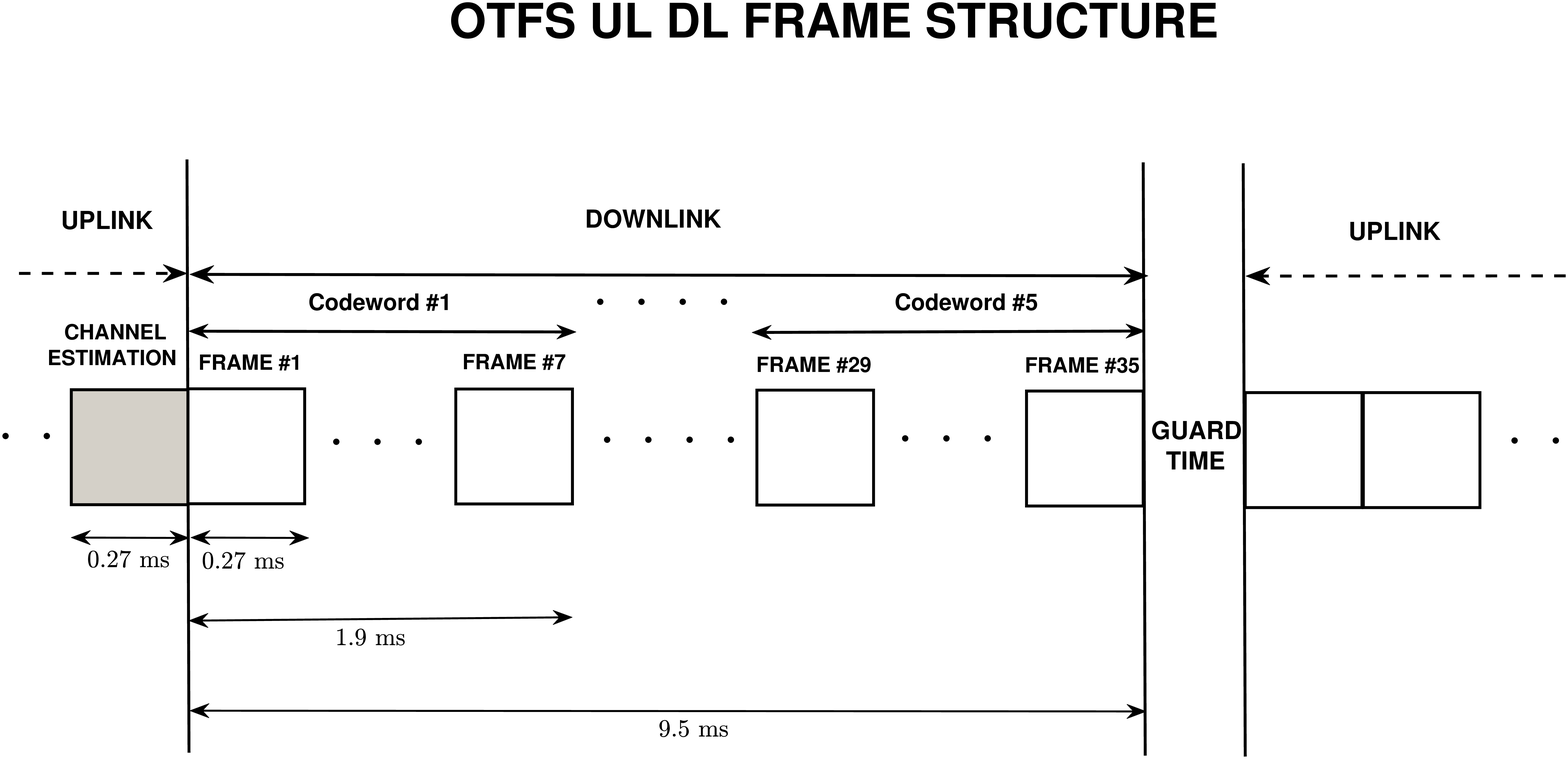}
\vspace{-0.1 cm}
\caption{OTFS Uplink (UL) and Downlink (DL) communication.} 
\vspace{-0.1cm}
\label{otfsuldl_fig}
\end{figure}

For the OTFS based system, since $T = 1/\Delta f = 66.66 \mu s$, each OTFS frame is $(\tau_{max} + NT) = 4.7 + 4 \times 66.66 =  271.37 \, \mu s$ in duration.
As shown in Fig.~\ref{otfsuldl_fig}, downlink communication consists of a sequence of $35$ consecutive OTFS frame of total duration $35 \times 271.37 \mu s = 9.5 $ ms. We consider a rate-$1/3$ Turbo code (see 3GPP Technical Standard (TS) 36.212: Multiplexing and Channel Coding) and $4$-QAM modulation of coded bits to information symbols.
Each codeword has $6144$ information bits and therefore consists of $3 \times (6144 + 4) / 2 =  9222$ $4$-QAM symbols. Since each OTFS frame can carry $M \times N = 330 \times 4 = 1320$ symbols,
a codeword spans $\lceil 9222/1320 \rceil = 7$ OTFS frames having duration $7 \times 271.37 \, \mu s = 1.9$ ms. 
Therefore, five codewords can be transmitted
in $35$ OTFS frames over a duration of $9.5$ ms. As discussed earlier in Section \ref{sysmodelsec}, the channel path delays, Doppler shifts and channel path gains are assumed to be fixed for a few tens of ms, and therefore in our simulations we consider them to be constant for one OTFS uplink phase followed by a downlink phase i.e., $2 \times  9.5 = 19$ ms (we assume alternate uplink and downlink communication phases, with each phase of $9.5$ ms duration).
Uplink channel estimation is performed in the last OTFS frame of each uplink communication phase (see Fig.~\ref{otfsuldl_fig}).

Although the OTFS channel matrices ($MN \times MN$) are large for large $M$ and/or $N$, they depend
on the path parameters (i.e., path gains, path delays and Doppler shifts) of only a small number of paths ($L_s \ll MN, s=1,2,\cdots,K$, see also (\ref{eq_channel})).  
Therefore, we consider low-complexity estimation of these parameters, which has been described in detail in Appendix \ref{appendixE}.
As discussed above, we consider the use of only one OTFS frame for transmission of uplink pilots. Through simulations (discussed later in this section) we show that the coded error rate performance with the proposed channel estimation is almost the same as that with perfect channel knowledge.
As the same channel estimate is used for both uplink and downlink communication and only one pilot OTFS frame is sufficient, the
channel estimation overhead is $\frac{100}{(2 \times 35)} = 1.43$ percent only.
\begin{figure}[t]
\vspace{-0.1 cm}
\hspace{-0.2 in}
\centering
\includegraphics[width= 3.1 in, height= 2.2 in]{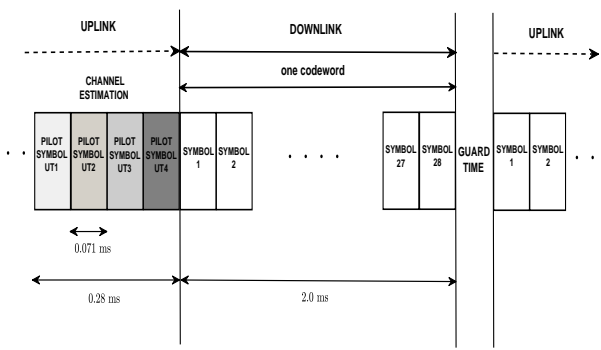}
\vspace{-0.1 cm}
\caption{OFDM UL and DL frames with one codeword transmitted in one DL frame.}
\vspace{-0.1cm}
\label{ofdmuldl_fig1}
\end{figure}

For the OFDM based massive MIMO system also, we consider alternate uplink and downlink communication phases of equal duration. Uplink pilots transmitted
by the UTs are used by the BS to acquire channel estimates. As shown in Fig.~\ref{ofdmuldl_fig1},
at the end of the uplink communication phase an OFDM symbol is dedicated for the transmission
of uplink pilot for each UT. Therefore with $K=4$ UTs, four OFDM symbols are dedicated for uplink channel estimation.
In each OFDM pilot symbol, pilots are transmitted on each sub-carrier and separate channel estimation for each sub-carrier
is performed on all the $Q$ receive antennas at the BS. In OFDM, the TF domain channel changes due to the
Doppler shift induced by each multi-path and therefore we consider two different UL-DL frame options. In option-I shown
in Fig.~\ref{ofdmuldl_fig1}, one codeword (same as that described earlier for the OTFS based system)
consisting of $9222$ $4$-QAM symbols is transmitted in the downlink in $\lceil 9222/330 \rceil =   28$ OFDM symbols, which
spans a duration of $28 \times 71.36 \, \mu s  =   2.0$ ms (as each OFDM symbol is $1/\Delta f + 4.7 \mu s = 71.36 \, \mu s$).
The channel estimation overhead with option-I ($K=4$ UTs) is therefore $\frac{K}{(2 \times 28)} \times 100 =  7.14$ percent.

\begin{figure}[t]
\vspace{-0.2 cm}
\hspace{-0.2 in}
\centering
\includegraphics[width= 3.1 in, height= 2.2 in]{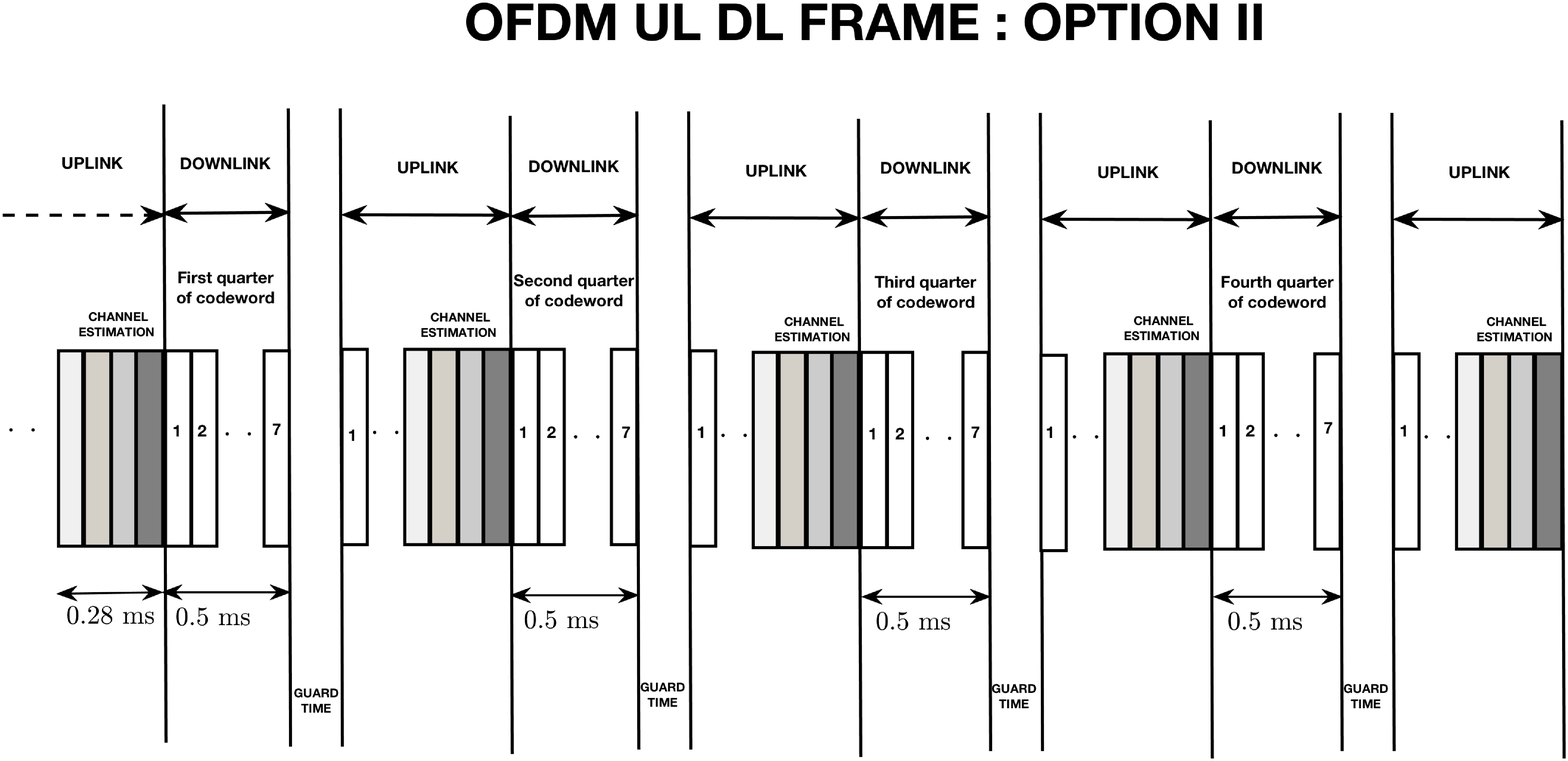}
\vspace{-0.1 cm}
\caption{OFDM UL and DL frames with one codeword spanning four DL frames.} 
\vspace{-0.1 cm}
\label{ofdmuldl_fig3}
\end{figure}    
In option-II as shown in Fig.~\ref{ofdmuldl_fig3}, the pilots are transmitted more frequently
to support higher mobility than that supported by option-I. Specifically, one codeword
is transmitted in four downlink communication phases. As channel estimation is performed at the end of each uplink phase, the channel estimation overhead with option-II
is therefore $\frac{K}{2 \times 7} \times 100 = 28.6$ percent.
%The guard time during the transition from downlink
%to uplink communication phase is small compared to the duration of the uplink and downlink communication phase and
%therefore we do not take it into account while calculating the channel estimation overhead.
Also,
the total time duration between the first channel estimation just before the start of transmission of a codeword and the end of transmission of that codeword is   
roughly $32 \times 71.36 \, \mu s = 2.28 $ ms and $53 \times 71.36 \, \mu s = 3.78$ ms for option-I and option-II respectively.
For these durations, the underlying channel path gains $h_{q,s,i}$, path delays $\tau_{s,i}$ and Doppler shifts $\nu_{s,i}$ do not change, but the TF domain channel changes due to the Doppler shifts of the channel paths. 
\begin{figure}[t]
\vspace{-0.1 cm}
\hspace{-0.2 in}
\centering
\includegraphics[width= 3.1 in, height= 2.2 in]{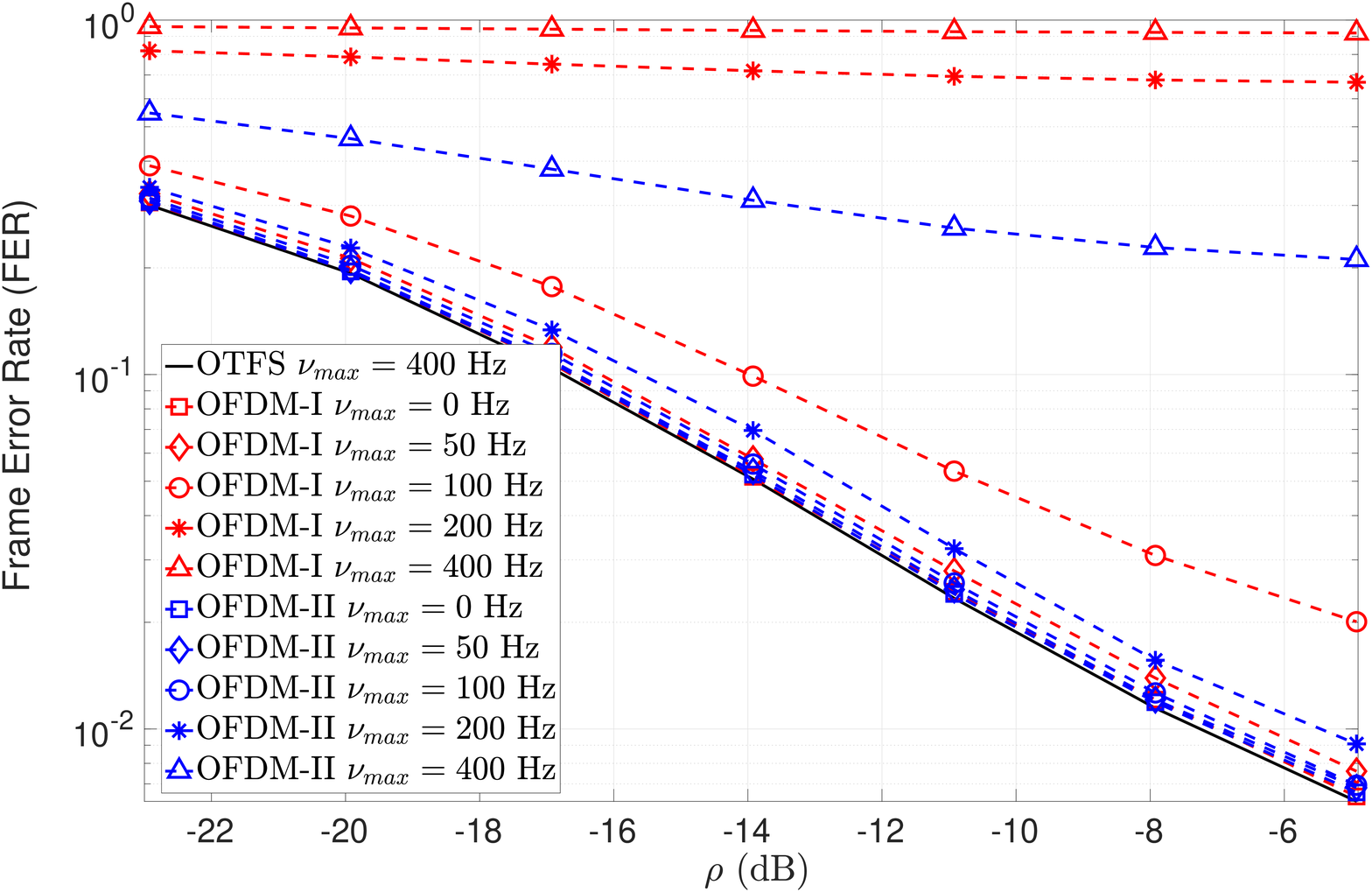}
\vspace{-0.1 cm}
\caption{Frame Error Rate vs. $\rho$ for OTFS and OFDM massive MIMO.} 
\vspace{-0.15cm}
\label{ofdmotfsfer_csi}
\end{figure}    
\begin{figure}[t]
\vspace{-0.1 cm}
\hspace{-0.2 in}
\centering
\includegraphics[width= 3.1 in, height= 2.0 in]{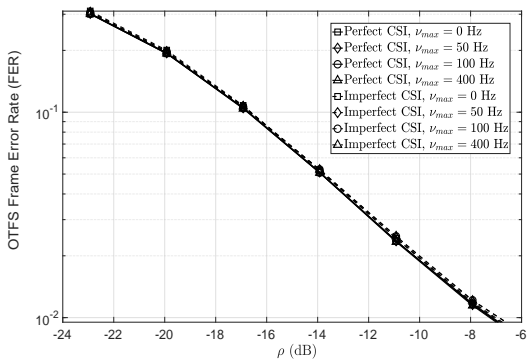}
\vspace{-0.1 cm}
\caption{FER vs. $\rho$ for OTFS massive MIMO.} 
\vspace{-0.1cm}
\label{otfsfer_csi}
\end{figure}    

In Fig.~\ref{ofdmotfsfer_csi}, we compare the Frame Error Rate (FER) (i.e., codeword error rate) performance of the proposed OTFS massive MIMO precoder (with the LCD detector)
and OFDM massive MIMO as a function of
increasing $\rho$ for a fixed $Q_h = Q_v = 14$ and $K=4$ UTs. We consider the imperfect CSI scenario
where the received pilot signal to noise ratio at each BS antenna is $26$ dB for both OTFS and OFDM systems.
We note that the FER performance of OFDM massive MIMO with option-I and option-II UL-DL frames
(denoted in the figure as OFDM-I and OFDM-II respectively), degrades with increase in Doppler spread.
Even with frequent transmission of uplink pilots (i.e., OFDM-II with a pilot overhead of $28.6$ percent),
the FER performance at $\nu_{max} = 400$ Hz is poor when compared to that of OTFS massive MIMO with a pilot overhead of
{\em only} $1.43$ percent. Therefore, the proposed OTFS massive MIMO precoder        
achieves significantly {\em better} error rate performance than OFDM massive MIMO, with much {\em lower} pilot overhead.

In Fig.~\ref{otfsfer_csi}, we plot the FER 
for OTFS massive MIMO with the proposed precoder (and the LCD detector) as a function of
increasing $\rho$ for a fixed number of BS antennas $(Q_h = Q_v = 14)$ and $K=4$ UTs.
We consider both the perfect and imperfect CSI scenarios. For the imperfect
CSI scenario, the received pilot signal to noise ratio at each BS antenna is $\rho_p = 26$ dB
(see Appendix \ref{appendixE}). From Fig.~\ref{otfsfer_csi} it is clear that the FER performance
for the perfect and imperfect CSI scenarios are the same, which shows the effectiveness of the
OTFS channel estimation considered in Appendix \ref{appendixE}. We also observe that
the FER performance is almost invariant of the Doppler spread.

\section{Conclusion}
In this paper we have proposed a low complexity OTFS multi-user massive MIMO precoder where
multi-user precoding is performed in the DD domain. As the optimal MMSE-SIC based detector at each UT has
prohibitive complexity when $(M,N)$ is large, we also propose a low complexity LCD
detector for detection of information symbols at each UT. Through analysis we show that
with increasing number of BS antennas the sum SE performance of the proposed LCD detector converges to that achieved by the 
optimal detector. Through simulations we also show that the proposed
OTFS multi-user massive MIMO precoder achieves significantly better SE than a OFDM multi-user massive MIMO precoder in high mobility scenarios
where the Doppler shift of the channel paths is high. Simulations of the coded error rate performance
of the proposed OTFS massive MIMO precoder and OFDM massive MIMO for the imperfect CSI scenario reveals that  
OTFS massive MIMO achieves significantly better FER than OFDM massive MIMO with much smaller pilot overhead. 

\appendices

\section{Derivation of (\ref{eqn10}) and (\ref{eq_channel})}
\label{appendixRavi}
Substituting the expression for $x_q(t)$ from (\ref{eqn6}) into the R.H.S. of (\ref{eqn4}), we get an expression for $y_s(t)$ in terms of the DD domain signals
$x_q[k,l], q=1,2,\cdots,Q$. Using this expression of $y_s(t)$ into the R.H.S. of (\ref{eqn7}) we get an expression for $Y_s[n,m]$ which when substituted in the R.H.S. of
(\ref{eqn8}) gives an expression for ${\widehat x_s}[k,l]$ in terms of $x_q[k,l]$, i.e.

{\vspace{-4mm}
\small
\begin{eqnarray}
\label{appRavieqn1}
\hspace{-3mm} {\widehat x_s}[k,l] & \hspace{-3mm}  \mya &  \hspace{-3mm} \frac{1}{MN} \sum\limits_{q=1}^Q \sum\limits_{k^{\prime}=0}^{N-1}\sum\limits_{l^{\prime}=0}^{M-1} \hspace{-1mm} x_q[k^{\prime},l^{\prime}]  \, {\Tilde h}_{q,s,k,l}[k^{\prime}, l^{\prime}]   + \, w_s[k,l] \nonumber \\
 {\Tilde h}_{q,s,k,l}[k^{\prime}, l^{\prime}] & \hspace{-3mm}  \Define &  \hspace{-3mm} \frac{1}{MN} \left(  h^{\mbox{\tiny{ici}}}_{q,s,k,l}[k^{\prime},l^{\prime}]  \, +   e^{-j 2 \pi \frac{k^{\prime}}{N}} h^{\mbox{\tiny{isi}}}_{q,s,k,l}[k^{\prime},l^{\prime}]  \right)
\end{eqnarray}
\normalsize}
where step (a) follows from equations $(53)$ and $(60)$ in Appendix C of \cite{channel} and the expressions for $h^{\mbox{\tiny{ici}}}_{q,s,k,l}[k^{\prime},l^{\prime}]$ and $h^{\mbox{\tiny{isi}}}_{q,s,k,l}[k^{\prime},l^{\prime}]$
follow from equations $(56)$ and $(61)$ respectively in \cite{channel} (see (\ref{eqnicisi}) in top of next page).
\begin{figure*}
{\vspace{-9mm}
\small
\begin{eqnarray}
\label{eqnicisi}
h^{\mbox{\tiny{ici}}}_{q,s,k,l}[k^{\prime},l^{\prime}] & \hspace{-3mm} = & \hspace{-3mm}  \sum\limits_{i=1}^{L_s} \frac{h_{q,s,i}}{M}  \left\{  \left( \sum_{n=0}^{N-1}e^{j2\pi n\Big(\frac{\nu_{s,i}}{\Delta f}-\frac{(k - k^{\prime})}{N}\Big)}  \right) \Biggr[   \sum_{p=0}^{M-1- l_{\tau_{s,i}}} \hspace{-5mm} e^{j2\pi\frac{p}{M}(\frac{\nu_{s,i}}{\Delta f})}\sum_{m=0}^{M-1}e^{-j2\pi\big(p+ l_{\tau_{s,i}} -l\big)\big(\frac{m}{M}\big)}  \sum_{m^{\prime}=0}^{M-1}e^{j2\pi(p-l^{\prime})\frac{m^{\prime}}{M}}  \Biggr] \right\} \nonumber \\
\hspace{-2mm} h^{\mbox{\tiny{isi}}}_{q,s,k,l}[k^{\prime},l^{\prime}] & \hspace{-3mm} = & \hspace{-3mm}  \sum\limits_{i=1}^{L_s} \frac{h_{q,s,i}}{M}  \left\{ \left( \sum_{n=0}^{N-1}e^{j2\pi n\Big(\frac{\nu_{s,i}}{\Delta f}-\frac{(k-k^{\prime})}{N}\Big)}  \right) \Biggr[    \sum_{p=M - l_{\tau_{s,i}}}^{M-1} \hspace{-5mm} e^{j2\pi\big(\frac{p-M}{M}\big)\frac{\nu_{s,i}}{\Delta f}}\sum_{m=0}^{M-1} \hspace{-1mm} e^{-j2\pi\big(p+ l_{\tau_{s,i}} -l\big)\big(\frac{m}{M}\big)}
 \sum_{m^{\prime}=0}^{M-1} \hspace{-1mm} e^{j2\pi(p-l^{\prime})\frac{m^{\prime}}{M}} \Biggr] \right\}.
 \end{eqnarray}
 \vspace{-3mm}
\begin{eqnarray*}
\hline 
\end{eqnarray*}
\vspace{-9mm}
 \normalsize}
\end{figure*}
Note that in (\ref{eqnicisi}), $l_{\tau_{s,i}} \Define \tau_{s,i} M \Delta f$ is an integer.
In the R.H.S. of equation $(61)$ in \cite{channel}, the expression for the ISI term consists of a summation over $n=1$ to $n= N-1$
whereas the expression for the ICI term in equation $(56)$ of \cite{channel} consists of a summation over $n=0$ to $n=N-1$. This is due to the fact that in \cite{channel},
instead of the CP, zero prefix of duration $\tau_{max}$ is used to avoid interference between consecutive OTFS frames in the time-domain. 
Due to this, in \cite{channel}, the sum over $n$ in the ISI term is $-1/N$ less than the sum over $n$ in the ICI term. This difference is significant when $N$ is not large
and can reduce the energy of the effective DD domain channel gains ${\Tilde h}_{q,s,k,l}[k^{\prime}, l^{\prime}]$. Therefore, in this paper we have considered
a CP of length $\tau_{max}$, due to which the summation over $n$ will be from $n=0$ to $n= N-1$ in both the ICI and the ISI terms (see (\ref{eqnicisi})).      

In (\ref{appRavieqn1}), the received additive noise sample on the $(k,l)$-th DDRE is given by 
\begin{eqnarray}
\label{noisekleqn}
w_s[k,l] & \hspace{-3mm}  \Define &  \hspace{-3mm}   \frac{1}{\sqrt{MN}} \sum\limits_{n=0}^{N-1}\sum\limits_{m=0}^{M-1}  W_s[n,m] \, e^{-j 2 \pi \left( \frac{n k}{N} - \frac{m l}{M}  \right)} \nonumber \\
& & \hspace{-3mm}  k=0,1,\cdots,N-1 \,,\, l=0,1,\cdots, M-1
\end{eqnarray}
where $W_s[n,m]$ is the received noise sample in the $(n,m)$-th TFRE.
Further, from (\ref{eqn4}) and (\ref{eqn7}), $W_s[n,m]$ is given by
\begin{eqnarray}
\label{noisenmeqn}
W_s[n,m] & \hspace{-3mm} = &  \hspace{-3mm}  \int g_{rx}^*(t - nT) {\Tilde w_s}(t) \, e^{-j 2 \pi m \Delta f (t - nT)} \, dt.
\end{eqnarray}
Since ${\Tilde w_s}(t)$ is AWGN with power spectral density $N_0$, from (\ref{noisekleqn}) and (\ref{noisenmeqn}) it follows
that
\begin{eqnarray}
\label{wkleqn}
w_s[k,l]  \, \sim \,  \mbox{\small{i.i.d.}} \, {\mathcal C}{\mathcal N}(0, N_0).
\end{eqnarray}
Let ${\bf H}_{q,s} \in {\mathbb C}^{MN \times MN}$ denote a matrix whose element in its $(kM + l + 1)$-th row and $(k^{\prime}M + l^{\prime} + 1)$-th column is ${\Tilde h}_{q,s,k,l}[k^{\prime}, l^{\prime}]$ (see (\ref{appRavieqn1})). Then, arranging ${\widehat x_s}[k,l], x_q[k,l]$ and $w_s[k,l]$ respectively into the vectors ${\widehat {\bf x}_s}, {{\bf x}_q}$ and
${{\bf w}_s}$ (see (\ref{eqn9})) and using the definition of ${\bf H}_{q,s}$ from (\ref{appRavieqn1}), we get (\ref{eqn10}). 
Further, from (\ref{eqn10}) and the expressions for $h^{\mbox{\tiny{ici}}}_{q,s,k,l}[k^{\prime},l^{\prime}]$ and $h^{\mbox{\tiny{isi}}}_{q,s,k,l}[k^{\prime},l^{\prime}]$ in (\ref{eqnicisi}) it follows that
\begin{eqnarray}
\label{hsqklaeqn}
{\Tilde h}_{q,s,k,l}[k^{\prime}, l^{\prime}] &    \Define &     \sum\limits_{i=1}^{L_s} h_{q,s,i}  \, a_{s,i,k,l}[k^{\prime},l^{\prime}]
\end{eqnarray}
where $a_{s,i,k,l}[k^{\prime},l^{\prime}]$ is given by (\ref{asikleqn}) (see top of next page).
\begin{figure*}
{\vspace{-4mm}
\small
\begin{eqnarray}
\label{asikleqn}
a_{s,i,k,l}[k^{\prime},l^{\prime}] & \hspace{-3mm}  \Define & \hspace{-3mm}  \left( \frac{1}{N}\sum_{n=0}^{N-1}e^{j2\pi n\Big(\frac{\nu_{s,i}}{\Delta f}-\frac{(k - k^{\prime})}{N}\Big)}  \right) \Biggr[   \sum_{p=0}^{M-1- l_{\tau_{s,i}}} \hspace{-5mm} e^{j2\pi\frac{p}{M}(\frac{\nu_{s,i}}{\Delta f})}  \overbrace{ \left\{ \frac{1}{M} \sum_{m=0}^{M-1}e^{-j2\pi\big(p+ l_{\tau_{s,i}} -l\big)\big(\frac{m}{M}\big)} \right\}  }^{= \delta\left[[p + l_{\tau_{s,i}} -l ]_{_M}\right]}   \overbrace{\left\{ \frac{1}{M} \sum_{m^{\prime}=0}^{M-1}e^{j2\pi(p-l^{\prime})\frac{m^{\prime}}{M}} \right\}}^{= \delta\left[[p-l^{\prime}]_{_M}\right]}  \Biggr]  \nonumber \\
& \hspace{-9mm}  &  \hspace{-8mm} \, + \,    \left( \frac{1}{N} \sum_{n=0}^{N-1}e^{j2\pi n\Big(\frac{\nu_{s,i}}{\Delta f}-\frac{(k-k^{\prime})}{N}\Big)}  \right) \Biggr[    \sum_{p=M - l_{\tau_{s,i}}}^{M-1} \hspace{-5mm} e^{j2\pi\big(\frac{p-M}{M}\big)\frac{\nu_{s,i}}{\Delta f}}  \underbrace{\left\{  \frac{1}{M} \sum_{m=0}^{M-1}e^{-j2\pi\big(p+ l_{\tau_{s,i}} -l\big)\big(\frac{m}{M}\big)} \right\}}_{= \delta\left[[p + l_{\tau_{s,i}} -l ]_{_M}\right]}
\underbrace{\left\{ \frac{1}{M} \sum_{m^{\prime}=0}^{M-1}e^{j2\pi(p-l^{\prime})\frac{m^{\prime}}{M}} \right\}}_{= \delta\left[[p-l^{\prime}]_{_M}\right]} \Biggr]   \nonumber \\
& \hspace{-15mm}  = &  \hspace{-5mm} \begin{cases}
0    , &    l \ne [ l^{\prime} + l_{\tau_{s,i}} ]_{_M} \\
\frac{1}{N} \left[ \sum_{n=0}^{N-1}e^{j2\pi n\Big(\frac{\nu_{s,i}}{\Delta f}-\frac{(k - k^{\prime})}{N}\Big)}  \right]     e^{j 2 \pi \frac{l^{\prime}}{M} \frac{\nu_{s,i}}{\Delta f}}  ,    &   l  = [ l^{\prime} +   l_{\tau_{s,i}}  ]_{_M} \,\,,\,\, l^{\prime} \in [ 0 \,,\, M - 1 -  l_{\tau_{s,i}} ] \\
\frac{1}{N} \left[ \sum_{n=0}^{N-1}e^{j2\pi n\Big(\frac{\nu_{s,i}}{\Delta f}-\frac{(k-k^{\prime})}{N}\Big)} \right]   e^{j 2 \pi {\Big (} \frac{l^{\prime}}{M} \frac{\nu_{s,i}}{\Delta f} - \frac{k^{\prime}}{N} -\frac{\nu_{s,i}}{\Delta f} {\Big )}}  ,    &   l  = [ l^{\prime} + l_{\tau_{s,i}} ]_{_M} \,\,,\,\, l^{\prime} \in [ M  - l_{\tau_{s,i}}  \,,\, M - 1] \\
\end{cases} \nonumber \\
 & & k^{\prime} = 0,1,\cdots, N-1 \,\,,\,\, k =  0,1,\cdots, N-1 \,\,,\,\,  l^{\prime} = 0,1,\cdots, M-1 \,\,,\,\,  l = 0,1,\cdots, M-1.
\end{eqnarray}
\vspace{-3mm}
\begin{eqnarray*}
\hline
\end{eqnarray*}
\normalsize}
\vspace{-9mm}
\end{figure*}
In (\ref{asikleqn}), the discrete impulse sequence $\delta[p] (p \in {\mathbb Z})$ is one when $p=0$ and is zero otherwise. In the first equation in (\ref{asikleqn}), $a_{s,i,k,l}[k^{\prime},l^{\prime}]$ is a sum of two terms, the first term in the first line and the second term below it.
Since the delays are integer multiple of $1/(M \Delta f)$, $ l_{\tau_{s,i}} = M \Delta f \tau_{s,i}$ is an integer. Therefore, in the R.H.S. of the first equation in (\ref{asikleqn}), we observe that each term is non-zero only when the summation variable $p$ equals $l^{\prime}$ (see the last inner summation over $m'$ which equals $\delta\left[ [ p - l^{\prime} ]_{_M}\right]$).
Also, the inner summation over $m$ is non-zero only when $p = \left[ l - l_{\tau_{s,i}} \right]_{_M}$. From these two facts it follows that $a_{s,i,k,l}[k^{\prime},l^{\prime}]$ is non-zero only when $ \left[ l^{\prime}  +  l_{\tau_{s,i}} -l\right]_{_M}$ is zero, i.e., $l =   \left[ l^{\prime}  +  l_{\tau_{s,i}} \right]_{_M}$. In the first line of the first equation in (\ref{asikleqn}), the summation over $p$ is from $p=0$ to $p=M - 1 - l_{\tau_{s,i}}$ and since this term can only be non-zero if $p = l^{\prime}$, it follows that this term contributes to $a_{s,i,k,l}[k^{\prime},l^{\prime}]$ if and only if $l^{\prime} \in  [ 0 \,,\, M - 1 -  l_{\tau_{s,i}} ]$. Similarly, the term in the second line of (\ref{asikleqn}) contributes to $a_{s,i,k,l}[k^{\prime},l^{\prime}]$ if and only if $l^{\prime} \in [ M  - l_{\tau_{s,i}}  \,,\, M - 1]$. The second equation in (\ref{asikleqn}) then follows from these observations.     

Let ${\bf A}_{s,i} \in {\mathbb C}^{MN \times MN}$ denote the matrix whose element in its $(kM + l + 1)$-th row and $(k^{\prime}M + l^{\prime} + 1)$-th column is $a_{s,i,k,l}[k^{\prime},l^{\prime}]$.  
Since ${\Tilde h}_{q,s,k,l}[k^{\prime}, l^{\prime}]$ denotes the element of ${\bf H}_{q,s}$ in its $(kM + l + 1)$-th row and $(k^{\prime}M + l^{\prime} + 1)$-th column, (\ref{eq_channel}) follows from (\ref{hsqklaeqn}). Let ${\bf a}_{s,i,r} \in {\mathbb C}^{MN \times 1}$ denote the $r$-th column of ${\bf A}_{s,i}^H$, $r=1,2,\cdots,MN$. For any $s = 1,2,\cdots, K$, $i=1,2,\cdots, L_s$ and $r=1,2,\cdots, MN$, the second equation in (\ref{asikleqn}) gives the expression for the elements in the $r$-th row of ${\bf A}_{s,i}$. Therefore, for the $r = (kM + l+1)$-th row of ${\bf A}_{s,i}$ i.e., ${\bf a}_{s,i,r}^H$, we have $k= \lfloor r/M \rfloor$ and $l = [r ]_{_M}$. With this $(k,l)$ pair (which corresponds to the $r$-th row of ${\bf A}_{s,i}$), from (\ref{asikleqn}) we have   

{\vspace{-4mm}
\small
\begin{eqnarray}
\label{eqnasirnorm}
\Vert  {\bf a}_{s,i,r} \Vert^2 & = & \frac{1}{N^2}  \sum\limits_{k^{\prime} = 0}^{N-1} \left\vert  \sum\limits_{n=0}^{N-1}   e^{j 2 \pi n \left( \frac{\nu_{s,i}}{\Delta f} - \frac{(k - k^{\prime})}{N}  \right)} \right\vert^2  \nonumber \\
& \hspace{-12mm}  = &  \hspace{-9mm} \frac{1}{N^2} \sum\limits_{k^{\prime} = 0}^{N-1} \sum\limits_{n_1 = 0}^{N-1} \sum\limits_{n_2 = 0}^{N-1} e^{j 2 \pi (n_1 - n_2)  \left( \frac{\nu_{s,i}}{\Delta f} - \frac{(k - k^{\prime})}{N}  \right) } \nonumber \\
& \hspace{-36mm} = &  \hspace{-21mm} \frac{1}{N^2} \sum\limits_{n_1 = 0}^{N-1} \sum\limits_{n_2 = 0}^{N-1} e^{j2 \pi (n_1 - n_2) \left( \frac{\nu_{s,i}}{\Delta f} - \frac{k }{N}  \right)} \underbrace{\left[  \sum\limits_{k^{\prime} = 0}^{N-1} e^{j 2 \pi (n_1 - n_2)  \frac{k^{\prime}}{N} }  \right]}_{= N \, \delta[n_1 - n_2 ]} \nonumber \\
& = & 1.
\end{eqnarray}
\normalsize}
Similarly, from the second equation in (\ref{asikleqn}) it also follows that
for any $r_1,r_2 = 1,2,\cdots, MN$ and $r_1 \ne r_2$
\begin{eqnarray}
\label{asir1r2eqn}
{\bf a}_{s,i,r_1}^H {\bf a}_{s,i,r_2} & = & 0.
\end{eqnarray}
From (\ref{eqnasirnorm}) and (\ref{asir1r2eqn}) we then have
\begin{eqnarray}
\label{AsiHAsi}
{\bf A}_{s,i} {\bf A}_{s,i}^H & = & {\bf I} \,,\,
{\bf A}_{s,i}^H {\bf A}_{s,i}   =   {\bf I}.
\end{eqnarray}

\section{Expression for $\gamma_{s,s',r,p}/Q \,\,\,\, (s \ne s')$}
\label{egvvprime}
Let ${\bf a}_{s,i,r} \in {\mathbb C}^{MN \times 1}$ denote the $r$-th column of ${\bf A}_{s,i}^H$. Then, from (\ref{eqn19}) we have

{\vspace{-4mm}
\small
\begin{eqnarray}
\label{eqnvvprime1}
\frac{\gamma_{s,s',r,p}}{Q} & = & \sum\limits_{q=1}^Q \sum\limits_{i=1}^{L_s} \sum\limits_{k=1}^{L_{s'}} \frac{h_{q,s,i} \, h_{q,s',k}^*}{Q} \, {\bf a}_{s,i,r}^H  \, {\bf a}_{s',k,p} \nonumber \\
& \hspace{-28mm} = &  \hspace{-16mm} \sum\limits_{i=1}^{L_s} \sum\limits_{k=1}^{L_{s'}} \frac{\sum\limits_{q=1}^Q h_{q,s,i} \, h_{q,s',k}^*}{Q} \, {\bf a}_{s,i,r}^H  \, {\bf a}_{s',k,p}.
\end{eqnarray}
\normalsize}
From (\ref{eqn3}) it follows that 
{\vspace{-2mm}
\small
\begin{eqnarray}
\label{eqnvvp12} 
\hspace{4mm} \frac{\sum\limits_{q=1}^Q  h_{q,s,i}    h_{q,s',k}^*}{Q}  &  &  \nonumber \\
& \hspace{-40mm} = &  \hspace{-22mm} \frac{g_{s,i} g^*_{s',k}}{Q_h Q_v} \sum\limits_{q=1}^Q \hspace{-1mm} e^{j 2 \pi \frac{d}{\lambda} \left( [ q-1]_{_{Q_h}} b_{s,s',i,k}  +  \left\lfloor  \frac{q-1}{Q_h} \right\rfloor  c_{s,s',i,k}  \right)} \nonumber \\
&  \hspace{-40mm}  = &  \hspace{-22mm} \frac{g_{s,i}  \, g^*_{s',k}}{Q_h Q_v} \left[  \hspace{-1mm} \sum\limits_{q_{_h} = 0}^{Q_h - 1} \hspace{-1mm} e^{j 2 \pi \frac{d}{\lambda} q_{_h} b_{s,s',i,k}} \right] \left[   \hspace{-1mm} \sum\limits_{q_{_v} = 0}^{Q_v - 1}  \hspace{-1mm} e^{j 2 \pi \frac{d}{\lambda} q_{_v} c_{s,s',i,k}} \right]  \nonumber \\
&  \hspace{-40mm}  = & \hspace{-22mm}  {\Bigg [} g_{s,i}  \, g^*_{s',k}  \, e^{j \pi \frac{d}{\lambda} \left( (Q_h - 1) b_{s,s',i,k} + (Q_v - 1) c_{s,s',i,k} \right)} \,   \nonumber \\
& & \frac{\sinc( \frac{d}{\lambda} b_{s,s',i,k} Q_h)}{\sinc( \frac{d}{\lambda} b_{s,s',i,k})} \frac{\sinc( \frac{d}{\lambda} c_{s,s',i,k} Q_v)}{\sinc( \frac{d}{\lambda} c_{s,s',i,k})} {\Bigg ]} \nonumber \\
&  & \hspace{-22mm} b_{s,s',i,k}  \Define  \sin(\phi_{s,i}) \sin(\theta_{s,i}) - \sin(\phi_{s',k}) \sin(\theta_{s',k}) \,\,, \nonumber \\
&  &  \hspace{-22mm} c_{s,s',i,k}  \Define   \cos(\theta_{s,i}) - \cos(\theta_{s',k}).
\end{eqnarray}
\normalsize
}
Using (\ref{eqnvvp12}) in (\ref{eqnvvprime1}) we get 

{\vspace{-4mm}
\small
\begin{eqnarray}
\label{eqnvvp34} 
\hspace{1mm} \frac{\gamma_{s,s',r,p}}{Q} & \hspace{-3.75mm}  =   &  \hspace{-4mm}  \sum\limits_{i=1}^{L_s} \sum\limits_{k=1}^{L_{s'}}  \hspace{-1mm} {\bf a}_{s,i,r}^H \, {\bf a}_{s',k,p} e^{j \pi \frac{d}{\lambda} \left( (Q_h - 1) b_{s,s',i,k} + (Q_v - 1) c_{s,s',i,k} \right)} \nonumber \\
&   &    \hspace{1mm}  g_{s,i} \, g_{s',k}^*  \frac{\sinc( \frac{d}{\lambda} b_{s,s',i,k} Q_h)}{\sinc( \frac{d}{\lambda} b_{s,s',i,k})} \frac{\sinc( \frac{d}{\lambda} c_{s,s',i,k} Q_v)}{\sinc( \frac{d}{\lambda} c_{s,s',i,k})}. 
\end{eqnarray}
\normalsize
}
Since the azimuthal and zenith angles for any two UTs are not the same, for any $s \ne s'$, at least one among $b_{s,s',i,k}$ and $c_{s,s',i,k}$ is not
zero. Hence, for all $s \ne s'$, $\frac{\sinc( \frac{d}{\lambda} b_{s,s',i,k} Q_h)}{\sinc( \frac{d}{\lambda} b_{s,s',i,k})} \frac{\sinc( \frac{d}{\lambda} c_{s,s',i,k} Q_v)}{\sinc( \frac{d}{\lambda} c_{s,s',i,k})} \rightarrow 0$ as $(Q_h,Q_v) \rightarrow \infty$. Using this fact in (\ref{eqnvvprime1}), for large $(Q_h,Q_v)$ we have
\begin{eqnarray}
\frac{\gamma_{s,s',r,p}}{Q} & \approx & 0, \,\,\,\, \mbox{\small{Large}} \,\,(Q_h,Q_v). 
\end{eqnarray}

\section{Expression for $ {\gamma_{s,s,r,r}}/{Q}$ }
\label{appendixA}
From (\ref{eqn19}) we have

{\vspace{-4mm}
\small
\begin{eqnarray}
\label{eqnA1}
\frac{\gamma_{s,s,r,r}}{Q} & \hspace{-3.5mm} = &  \hspace{-3.5mm} \sum_{q=1}^{Q} \sum_{i=1}^{L_s}  \hspace{-1mm} \left[ \frac{|h_{q,s,i}|^{2}}{Q}  \Vert {\bf a}_{s,i,r} \Vert^2 + \hspace{-1mm} \sum\limits_{k=1 \atop k \neq i}^{L_s} \hspace{-1mm} \frac{h_{q,s,i}  h^*_{q,s,k}}{Q}   {\bf a}_{s,i,r}^H {\bf a}_{s,k,r} \right] \nonumber \\
\hspace{-8.5mm}  & \hspace{-17.5mm}  \mya & \hspace{-8.5mm}  \sum\limits_{i=1}^{L_s} \vert g_{s,i} \vert^2   \nonumber \\
& &  \hspace{-13mm} + \sum\limits_{i=1}^{L_s}\sum\limits_{k=1 \atop k \neq i}^{L_s}  \hspace{-1mm} {\bf a}_{s,i,r}^H {\bf a}_{s,k,r}   {\Bigg \{} e^{j \pi \frac{d}{\lambda} \left( (Q_h - 1) b_{s,s,i,k} + (Q_v - 1) c_{s,s,i,k} \right)} \nonumber \\
& &   \hspace{5.5mm} g_{s,i} g_{s,k}^*   \frac{\sinc( \frac{d}{\lambda} b_{s,s,i,k} Q_h)}{\sinc( \frac{d}{\lambda} b_{s,s,i,k})} \frac{\sinc( \frac{d}{\lambda} c_{s,s,i,k} Q_v)}{\sinc( \frac{d}{\lambda} c_{s,s,i,k})}  {\Bigg \}}
\end{eqnarray}
\normalsize}
where step (a) follows from (\ref{eqn3}) and $b_{s,s,i,k},c_{s,s,i,k}$ are given by (\ref{eqnvvp12}) with $s' = s$. In step (a) we have also used the fact that 
$\Vert {\bf a}_{s,i,r} \Vert^2 = 1$ (see (\ref{eqnasirnorm})). Further, since the azimuthal and zenith angles for any two channel paths to a UT are not the same, for any $i \ne k$, at least one among $b_{s,s,i,k}$ or $c_{s,s,i,k}$ is not zero.
Hence, as $(Q_h,Q_v) \rightarrow \infty$, the term $ \frac{\sinc( \frac{d}{\lambda} b_{s,s,i,k} Q_h)}{\sinc( \frac{d}{\lambda} b_{s,s,i,k})} \frac{\sinc( \frac{d}{\lambda} c_{s,s,i,k} Q_v)}{\sinc( \frac{d}{\lambda} c_{s,s,i,k})} \rightarrow 0$
for all $i \ne k$. Using this fact in (\ref{eqnA1}), for large $(Q_h,Q_v)$ we have
\begin{eqnarray}
\frac{\gamma_{s,s,r,r}}{Q}  & \approx &   \sum\limits_{i=1}^{L_s} \vert g_{s,i} \vert^2 , \,\,\,\, \mbox{\small{Large}} \,\,(Q_h,Q_v). 
\end{eqnarray}   

\section{Expression for $\gamma_{s,s,r,p}/Q \,\,\,\, (r \ne p)$}
\label{appendixC}
From (\ref{eqn19}) we have

{\vspace{-4mm}
\small
\begin{eqnarray}
\label{eqnC1}
\frac{\gamma_{s,s,r,p}}{Q} & = &   \sum\limits_{i=1}^{L_s} {\Bigg (}  \frac{ \sum\limits_{q=1}^Q \vert h_{q,s,i} \vert^2}{Q} {\bf a}_{s,i,r}^H {\bf a}_{s,i,p}  \nonumber \\
& & \hspace{10mm} +  \sum\limits_{k=1 \atop k \ne i}^{L_s} \frac{ \sum\limits_{q=1}^Q  h_{q,s,i} h_{q,s,k}^*}{Q} {\bf a}_{s,i,r}^H {\bf a}_{s,k,p} {\Bigg)} \nonumber \\
& \hspace{-20mm} \mya & \hspace{-12mm}  \sum\limits_{i=1}^{L_s}\sum\limits_{k=1 \atop k \neq i}^{L_s}  \hspace{-1mm} {\bf a}_{s,i,r}^H {\bf a}_{s,k,p}   {\Bigg [} e^{j \pi \frac{d}{\lambda} \left( (Q_h - 1) b_{s,s,i,k} + (Q_v - 1) c_{s,s,i,k} \right)} \nonumber \\
& &     \hspace{-4mm}  g_{s,i} \, g_{s,k}^* \, \frac{\sinc( \frac{d}{\lambda} b_{s,s,i,k} Q_h)}{\sinc( \frac{d}{\lambda} b_{s,s,i,k})} \frac{\sinc( \frac{d}{\lambda} c_{s,s,i,k} Q_v)}{\sinc( \frac{d}{\lambda} c_{s,s,i,k})}  {\Bigg ]} 
\end{eqnarray}
\normalsize}
where step (a) follows from (\ref{eqn3}) and the fact that for any $p \ne r$, ${\bf a}_{s,i,r}^H {\bf a}_{s,i,p} = 0$ (see (\ref{asir1r2eqn})). Further, since the azimuthal and zenith angles are not the same for any two channel paths to a UT, for any $i \ne k$, at least one among $b_{s,s,i,k}$ or $c_{s,s,i,k}$ is not zero.
Hence, as $(Q_h,Q_v) \rightarrow \infty$, the term $ \frac{\sinc( \frac{d}{\lambda} b_{s,s,i,k} Q_h)}{\sinc( \frac{d}{\lambda} b_{s,s,i,k})} \frac{\sinc( \frac{d}{\lambda} c_{s,s,i,k} Q_v)}{\sinc( \frac{d}{\lambda} c_{s,s,i,k})} \rightarrow 0$
for all $i \ne k$. Using this fact in (\ref{eqnC1}), for large $(Q_h,Q_v)$ we have
\begin{eqnarray}
\frac{\gamma_{s,s,r,p}}{Q} & \approx &    0, \,\,\,\, \mbox{\small{Large}} \,\,(Q_h,Q_v). 
\end{eqnarray}

\section{Proof of Theorem \ref{thm34}}
\label{appendixD}
We firstly find a good approximation to $C({\mathcal P})$ in the large $(Q_h,Q_v)$ regime with constant $\rho Q$.
From (\ref{eqn19}) we get

{\vspace{-4mm}
\small
\begin{eqnarray}
\label{Ginteqn}
\frac{{\bf G}_{s,s}}{Q} & \hspace{-3mm} = &  \hspace{-3mm}  \sum_{q=1}^{Q} \sum_{i=1}^{L_s}\left( \frac{|h_{q,s,i}|^{2}}{Q} \mathbf{A}_{s,i} \mathbf{A}^H_{s,i}+\sum_{k=1 \atop k \neq i}^{L_s} \frac{h_{q,s,i}  h^*_{q,s,k}}{Q} \mathbf{A}_{s,i} \mathbf{A}^H_{s,k}\right) \nonumber \\
& \hspace{-16mm}  \mya &  \hspace{-10mm}   \sum\limits_{i=1}^{L_s}   \, \frac{\sum\limits_{q=1}^Q \vert h_{q,s,i} \vert^2}{Q}  {\bf I}   \, + \,  \sum\limits_{i=1}^{L_s} \sum\limits_{k=1 \atop k \neq i}^{L_s}   \frac{\sum\limits_{q=1}^Q h_{q,s,i}  h^*_{q,s,k}}{Q}  \mathbf{A}_{s,i} \mathbf{A}^H_{s,k}  \nonumber \\
& \hspace{-16mm}  \myapproxb & \hspace{-10mm}  \left( \sum\limits_{i=1}^{L_s} \vert g_{s,i}  \vert^2 \right) {\bf I}  \,\, , \,\, \mbox{\small{Large}} \,\,(Q_h,Q_v)
\end{eqnarray}
\normalsize}
where step (a) follows from the expression for $h_{q,s,i}$ in (\ref{eqn3}) and the fact that ${\bf A}_{s,i} {\bf A}_{s,i}^H = {\bf I}$ (see  (\ref{AsiHAsi})).
Step (b) follows from the fact that, for all $i \ne k$, $\left( {\sum\limits_{q=1}^Q h_{q,s,i}  h^*_{q,s,k}}/{Q}\right)  \rightarrow 0 $ as $(Q_h,Q_v) \rightarrow \infty$ (see (\ref{eqnC1}) and the paragraph following it).
Similarly, from (\ref{eqn19}) we also have 

{\vspace{-4mm}
\small
\begin{eqnarray}
\label{Ginteqn1}
\frac{{\bf G}_{s,s'}}{Q} & = & \sum\limits_{i=1}^{L_s} \sum\limits_{k=1}^{L_{s'}}  {\Bigg (} \frac{\sum\limits_{q=1}^Q h_{q,s,i} h_{q,s',k}^* }{Q}  {\Bigg )}  {\bf A}_{s,i} {\bf A}_{s',k}^H \nonumber \\
& \approx & {\bf 0}  , \,\, \mbox{\small{Large}} \,\,(Q_h,Q_v).
\end{eqnarray}
\normalsize}
where we have used the fact that $\left({\sum\limits_{q=1}^Q h_{q,s,i} h_{q,s',k}^* }/{Q} \right)  \rightarrow 0$ as $(Q_h,Q_v) \rightarrow \infty$ (see (\ref{eqnvvp12}) and the paragraph following (\ref{eqnvvp34})).
Using the large $(Q_h,Q_v)$ approximation for ${\bf G}_{s,s'}/Q$ from (\ref{Ginteqn1}) and the expression for $\eta$ from (\ref{eqn14}), we get

{\vspace{-4mm}
\small
\begin{eqnarray}
\label{eqnprfthm2e}
\frac{E_T}{\eta N_o}  \sum\limits_{s' = 1, \atop s' \ne s}^K {\bf G}_{s,s'} {\bf G}_{s,s'}^H & = & \frac{E_TQ}{\frac{\eta}{Q} N_o} \sum\limits_{s' = 1, \atop s' \ne s}^K \frac{{\bf G}_{s,s'}}{Q} \frac{{\bf G}^H_{s,s'}}{Q} \nonumber \\
& \hspace{-41mm}  \mya &  \hspace{-24mm} \frac{\rho Q}{\sum\limits_{{\Tilde s}=1}^K  \sum\limits_{i=1}^{L_{\Tilde s}}  \beta_{{\Tilde s},i} } \sum\limits_{s' = 1, \atop s' \ne s}^K \frac{{\bf G}_{s,s'}}{Q} \frac{{\bf G}^H_{s,s'}}{Q}  \nonumber \\
& \hspace{-20mm}  \approx &  \hspace{-11mm} {\bf 0}, \,\,\,  \mbox{\small{Large}} \,\,(Q_h,Q_v) \,\, \mbox{\small{and constant}} \,\, \rho Q
\end{eqnarray}
\normalsize}
where step (a) follows from (\ref{rhodefeqn}) and the last step follows from the fact that $\rho Q$ is kept constant with increasing $(Q_h,Q_v)$.
Using (\ref{Ginteqn}) and (\ref{eqnprfthm2e}) in (\ref{Cvdefeqn}) we get the following large $(Q_h,Q_v)$ with constant $\rho Q$ approximation for $C_s({\mathcal P})$ as   

{\vspace{-4mm}
\small
\begin{eqnarray}
\label{Cvapproxeqn}
C_s({\mathcal P}) & \myapproxa & \frac{1}{MN \left( 1 + \frac{\tau_{max}}{NT} \right)} \log_2 \left\vert {\bf I} + \frac{MN (\rho Q)}{\eta/Q} \frac{{\bf G}_{s,s}}{Q} \frac{{\bf G}_{s,s}^H}{Q} \right\vert   \nonumber \\
& \hspace{-10mm}  \myb & \hspace{-7mm} \frac{1}{MN\left( 1 + \frac{\tau_{max}}{NT} \right)} \sum\limits_{r=1}^{MN } \log_2 \left(1 + \frac{\rho Q}{\frac{\eta/Q}{MN}} \left( \sum\limits_{i=1}^{L_s} \vert g_{s,i} \vert^2  \right)^2 \right)  \nonumber \\
& \hspace{-10mm}  \myapproxc &  \hspace{-7mm} \frac{1}{MN \left( 1 + \frac{\tau_{max}}{NT} \right)}  \sum\limits_{r=1}^{MN} \log_2 \left(1 +  \frac{\rho Q \left( \sum\limits_{i=1}^{L_s} \vert g_{s,i} \vert^2  \right)^2 }{\sum\limits_{{\Tilde s} =1}^K\sum\limits_{i=1}^{L_{\Tilde s}} \beta_{{\Tilde s},i} }  \right)
\end{eqnarray}
\normalsize}
where step (a) follows from (\ref{rhodefeqn}) and (\ref{eqnprfthm2e}). Step (b) follows from (\ref{Ginteqn}) and
step (c) follows from the expression for $\eta$ in (\ref{eqn14}). From the large $Q$ (with constant $\rho \, Q$) approximation of $I_{s,r}$ in (\ref{eqn269}) of Corollary \ref{cor11} we have

{\vspace{-4mm}
\small
\begin{eqnarray}
\label{Rvapproxeqn}
R_s({\mathcal P}) & \mya & \frac{1}{MN\left( 1 + \frac{\tau_{max}}{NT} \right)} \sum\limits_{r=1}^{MN} I_{s,r}({\mathcal P})  \nonumber \\
& \hspace{-23mm}  \myapproxb &  \hspace{-14mm} \frac{1}{MN\left( 1 + \frac{\tau_{max}}{NT} \right)} \sum\limits_{r=1}^{MN}  \log_2 \left( 1 + \frac{\rho Q \left( \sum\limits_{i=1}^{L_s} \vert g_{s,i} \vert^2  \right)^2}{ \sum\limits_{{\Tilde s}=1}^K\sum\limits_{i=1}^{L_{\Tilde s}} \beta_{{\Tilde s},i} } \right)
\end{eqnarray} 
\normalsize}
where step (a) follows from the first definition in (\ref{eqnlem1}) and step (b) follows from (\ref{eqn269}). 
From (\ref{Rvapproxeqn}) and (\ref{Cvapproxeqn}) it follows that in the large $(Q_h,Q_v)$ regime (with constant $\rho Q$), $C_s({\mathcal P}) \approx R_s({\mathcal P})$. The proof then follows from using this fact in (\ref{speceffeqn}) and (\ref{rsumlcdeqn}). 

\section{OTFS Channel Estimation}
\label{appendixE}
\begin{figure}[t]
\vspace{-0.2 cm}
\hspace{-0.2 in}
\centering
\includegraphics[width= 2.7 in, height= 1.8 in]{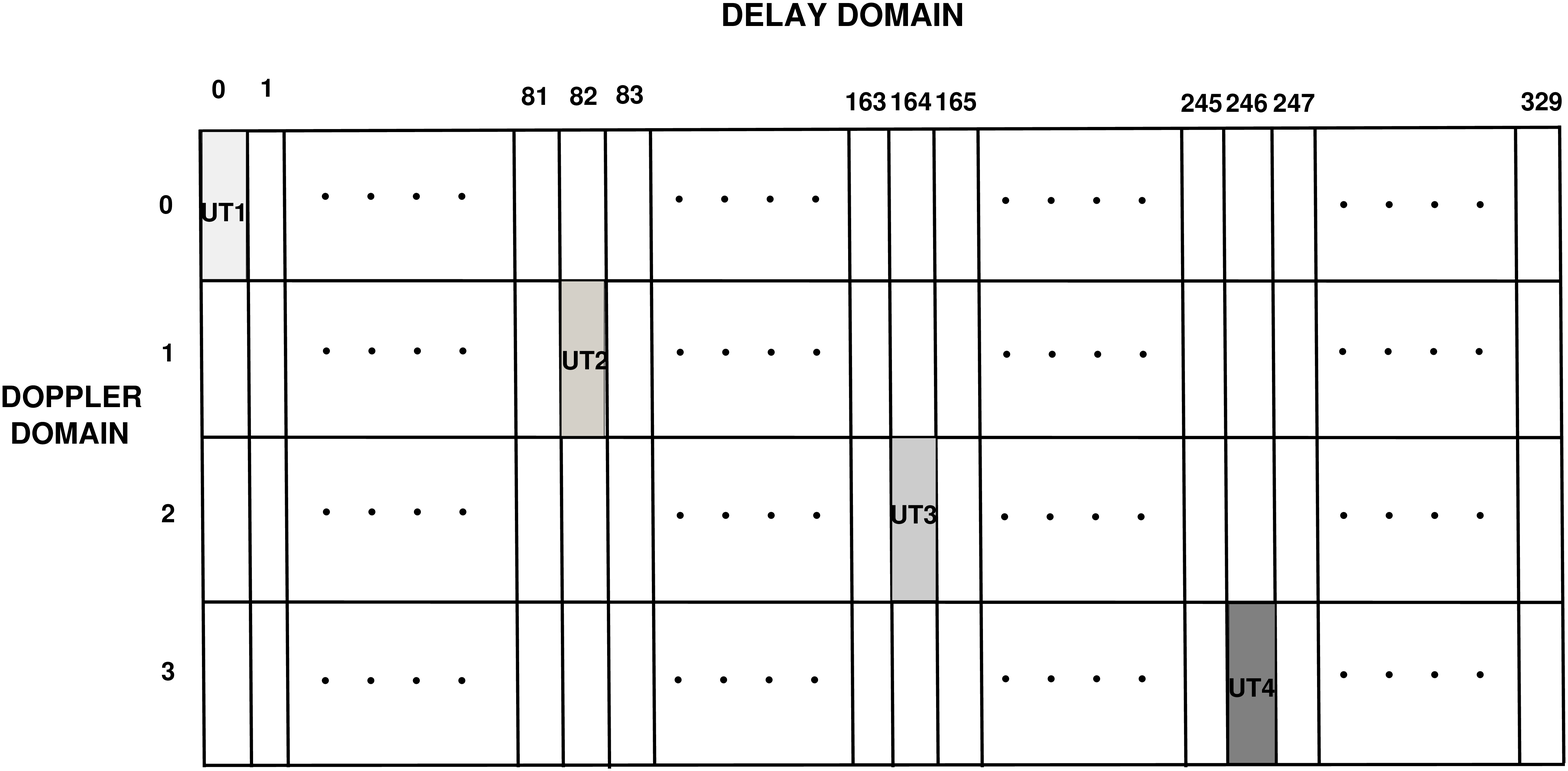}
\vspace{-0.05 cm}
\caption{Placement of uplink pilots in the DD domain. Pilots are transmitted
by the UTs on the shaded DDREs.} 
\vspace{-0.05cm}
\label{otfspilots_fig}
\end{figure}
In this section, we discuss OTFS channel estimation considered by us in the numerical section.
The OTFS massive MIMO channel estimation discussed here in the following is suited for systems where the BS does
not have accurate knowledge of the antenna array steering vector.
It is noted that prior works on OTFS massive MIMO channel estimation assume the exact knowledge of the antenna array steering vector
at the BS \cite{ChEstOTFS1, ChEstOTFS2, ChEstOTFS3}. 
In the considered TDD system, each UT sends a pilot symbol on a single designated DDRE in
the uplink OTFS frame dedicated for uplink pilot transmission. For example, the placement
of $K=4$ pilots for the $4$ UTs is shown in Fig.~\ref{otfspilots_fig} for $M = 330, N=4$.
In our simulations we have considered a maximum path delay of $\tau_{max} = 4.7 \, \mu s$,
due to which a pilot transmitted by a UT on the $(k,l) = (0,0)$-th DDRE will be received between the $l=0$-th and the
$l = \lceil \tau_{max} M \Delta f \rceil = 24$-th DDRE along the delay domain ($\Delta f = 1/T = 15$ KHz).
If we consider a maximum Doppler shift of $\nu_{max} = 1600$ Hz,
then the
same pilot will be spread over almost all the DDREs along the Doppler domain.

In order that the received pilots for the $K=4$ UTs  do not interfere with each other along the delay domain,
we place the pilots for the UTs at an interval of $\lfloor M/4 \rfloor = 82$ DDREs along the delay domain (see Fig.~\ref{otfspilots_fig}).
The pilots for the four UTs
are placed on DDREs having different Doppler domain indices in order that the interference between the received pilots is even smaller.
Let the pilot for the $s$-th UT be located at the $(k_s,l_s)$-th DDRE, where $k_s = (s-1), l_s = (s-1) \lfloor M/4 \rfloor, s=1,2,3,4$.
For the $s$-th UT we define its pilot region ${\mathcal R}_s = \left\{ (k,l) \, | \, k=k_s, l - l_s \in [ 0 \,,\, 82 ]   \right\}$ by the set of those DDREs where significant energy would be received only from
the pilot transmitted by the $s$-th UT.   

The DD domain pilot signal transmitted by the $s$-th UT is $x_s[k,l] = \sqrt{E_p}$ when $(k=k_s, l=l_s)$
and is zero for all other values of $(k,l)$. Here $E_p$ is the pilot energy.
We assume power control for the uplink pilots, such that the pilot energy received from each UT
is the same, i.e., $E_p$. 
The average uplink pilot power received from each UT is therefore $E_p/(NT)$.
The noise power at each BS antenna in the communication bandwidth of $M \Delta f$ Hz is $M \Delta f N_o$.
Therefore, the received pilot signal to noise ratio at the BS is $\rho_p \Define E_p/(NT M \Delta f N_o)  = E_p/(M N N_o)$ for each UT.

Let ${\widehat x_q}[k,l]$ denote the pilot signal received at the $q$-th BS antenna in the $(k,l)$-th DDRE, $k=0,1,\cdots, N-1$, $l=0,1,\cdots, M-1$.
Further, let ${\widehat {\bf x}_q} \in {\mathbb C}^{MN \times 1}$ denote the vector of all symbols received on the $MN$ DDREs at the $q$-th BS antenna,
which is given by (\ref{eqn701}) (see top of next page). 
From the system model described in Section \ref{sysmodelsec} it follows that the DD domain pilot signal received at the $q$-th BS antenna       
is given by

{\vspace{-4mm}
\small
\begin{eqnarray}
\label{eqn2121}
{\widehat {\bf x}_q} &  \hspace{-3mm} = &  \hspace{-3mm} \sum_{s=1}^K {\bf H}_{q,s} {\bf x}_s \, + \, {\bf w}_q  \, = \,  \sum_{s=1}^K \sum\limits_{i=1}^{L_s} h_{q,s,i}  \, {\bf A}_{s,i}  \, {\bf x}_s \, + \, {\bf w}_q
\end{eqnarray}
\normalsize}
%where ${\bf H}_{q,s}$ is the effective DD domain channel matrix between the $s$-th UT and the $q$-th BS antenna and is given by (\ref{eq_channel}).
where ${\bf x}_s$ is the vector of DD domain pilot symbols transmitted by the $s$-th UT and is given by (\ref{eqn701}) (see top of next page).
The AWGN in the $(k,l)$-th DDRE at the $q$-th BS antenna is denoted by $w_q[k,l], k=0,1,\cdots, N-1, l=0,1,\cdots, M-1$, and the vector of
DD domain AWGN samples is denoted by ${\bf w}_q$ (see (\ref{eqn701}) on top of next page).
\begin{figure*}
{\vspace{-9mm}
\small
\begin{eqnarray}
\label{eqn701}
{\widehat {\bf x}_q} &  \Define  &  \left(  {\widehat x_q}[0,0] , \cdots, {\widehat x_q}[0,M-1], {\widehat x_q}[1,0], \cdots, {\widehat x_q}[1,M-1], \cdots, {\widehat x_q}[N-1,0], \cdots,    {\widehat x_q}[N-1,M-1]\right)^T, \nonumber \\
{ {\bf w}_q} &  \Define  &  \left(  { w_q}[0,0] , \cdots, { w_q}[0,M-1], { w_q}[1,0], \cdots, { w_q}[1,M-1], \cdots, { w_q}[N-1,0], \cdots,    { w_q}[N-1,M-1]\right)^T, \nonumber \\
{ {\bf x}_s} &  \Define  &  \left(  { x_s}[0,0] , \cdots, { x_s}[0,M-1], { x_s}[1,0], \cdots, { x_s}[1,M-1], \cdots, { x_s}[N-1,0], \cdots,    { x_s}[N-1,M-1]\right)^T.
\end{eqnarray}
\normalsize
\vspace{-9mm}
}
\end{figure*}
Since, the $s$-th UT transmits an uplink pilot symbol only on the $(k_s,l_s)$-th DDRE, from (\ref{eqn2121}) we have

{\vspace{-6mm}
\small
\begin{eqnarray}
\label{eqn2122}
{\widehat {\bf x}_q} & = & \sqrt{E_p} \sum_{s=1}^K \sum\limits_{i=1}^{L_s} h_{q,s,i}  \, {\Tilde {\bf a}}_{s,i}(l_{\tau_{s,i}}, \nu_{s,i})  \, + \, {\bf w}_q
\end{eqnarray}
\normalsize}
where ${\Tilde {\bf a}}_{s,i}(l_{\tau_{s,i}}, \nu_{s,i}) \in {\mathbb C}^{MN \times 1}$ is the $(k_s M + l_s + 1)$-th column of the matrix ${\bf A}_{s,i}$. The $(KM + l+1)$-th element
of ${\Tilde {\bf a}}_{s,i}(l_{\tau_{s,i}}, \nu_{s,i})$ is $a_{s,i,k,l}[k_s, l_s]$ which is given by the expression for $a_{s,i,k,l}[k', l']$ in (\ref{asikleqn}) with $k^{\prime}= k_s$ and $l^{\prime} = l_s$ (in (\ref{asikleqn}) note the dependence on $l_{\tau_{s,i}}$ and $\nu_{s,i}$).
%From this expression in (\ref{asikleqn}) it is clear that ${\Tilde {\bf a}}_{s,i}$ depends only on the path delay $l_{\tau_{s,i}}$ and Doppler shift $\nu_{s,i}$ for the $i$-th path
%between the BS and the $s$-th UT.
%Since the DD domain AWGN samples at the BS antennas are i.i.d. complex Gaussian,
The maximum likelihood (ML) estimate of the channel path gains $\{ h_{q,s,i} \}_{q=1,s=1,i=1}^{Q,K, L_s}$, the path delays $\{ l_{\tau_{s,i}}   \}_{s=1,i=1}^{K, L_s}$ and the Doppler shifts $\{ {\nu_{s,i}}   \}_{s=1,i=1}^{K, L_s}$
is then given by (\ref{eqn2123}) (see top of next page).
\begin{figure*}
{\small
\vspace{-2mm}
\begin{eqnarray}
\label{eqn2123}
\left( \{ \widehat{h_{q,s,i}} \}_{q=1,s=1,i=1}^{Q,K, L_s}, \{ \widehat {l_{\tau_{s,i}}}   \}_{s=1,i=1}^{K, L_s} ,   \{ \widehat{\nu_{s,i}}   \}_{s=1,i=1}^{K, L_s}  \right) & \hspace{-2mm}  \Define &  \hspace{-2mm} \arg  \hspace{-10mm}  \min_{\substack{   \{ h_{q,s,i} \}_{q=1,s=1,i=1}^{Q,K, L_s}, \{ l_{\tau_{s,i}}   \}_{s=1,i=1}^{K, L_s}, \\ \{ {\nu_{s,i}}   \}_{s=1,i=1}^{K, L_s}} }  \sum_{q=1}^Q  \left\Vert  {\widehat {\bf x}_q}   -  \sqrt{E_p} \sum_{s=1}^K \sum\limits_{i=1}^{L_s} h_{q,s,i}  \, {\Tilde {\bf a}}_{s,i}(l_{\tau_{s,i}}, \nu_{s,i}) \right\Vert^2.
\end{eqnarray}
\normalsize}
\end{figure*}
The objective function in (\ref{eqn2123}) can be further expressed as in (\ref{eqn2223}) (see top of next page).
\begin{figure*}
{\small
\vspace{-9mm}
\begin{eqnarray}
\label{eqn2223}
 \frac{1}{Q} \sum_{q=1}^Q  \left\Vert  {\widehat {\bf x}_q}  \, -  \, \sqrt{E_p} \sum_{s=1}^K \sum\limits_{i=1}^{L_s} h_{q,s,i}  \, {\Tilde {\bf a}}_{s,i}(l_{\tau_{s,i}}, \nu_{s,i}) \right\Vert^2  & \hspace{-2mm}  = &  \hspace{-2mm} \frac{1}{Q} \sum_{q=1}^Q  \Vert   {\widehat {\bf x}_q}  \Vert^2  \, - \, \frac{2}{Q}  \sqrt{E_p} \sum\limits_{q=1}^Q \sum\limits_{s=1}^K \sum\limits_{i=1}^{L_s} \Re \left[ h_{q,s,i}  \,  {\widehat {\bf x}_q}^H  \, {\Tilde {\bf a}}_{s,i}(l_{\tau_{s,i}}, \nu_{s,i}) \right]  \nonumber \\
& &  \hspace{-8mm}  + \, E_p \sum\limits_{s=1}^K \sum\limits_{i=1}^{L_s}  \underbrace{{\Bigg (}  \frac{\sum\limits_{q=1}^Q \vert h_{q,s,i} \vert^2 }{Q}  {\Bigg )}}_{= \, \vert g_{s,i} \vert^2  \,  >   \,  0, \, \mbox{\small{see}} \, (\ref{eqn3}) } \, \Vert  {\Tilde {\bf a}}_{s,i}(l_{\tau_{s,i}}, \nu_{s,i}) \Vert^2  \nonumber \\
& &  \hspace{-8mm}  +  \,  E_p  \sum\limits_{s=1}^K \sum\limits_{i_1=1}^{L_s}  \sum\limits_{\substack{i_2=1 \\  i_2  \ne i_1}}^{L_s}  \underbrace{{\Bigg (}  \frac{\sum\limits_{q=1}^Q   h_{q,s,i_1}^* h_{q,s,i_2}   }{Q}  {\Bigg )}}_{\substack{\rightarrow 0, \, \mbox{\small{as}} \, (Q_h, Q_v) \rightarrow \infty \\ \mbox{\small{see}}  \, (\ref{eqnC1}) \, \mbox{\small{and para. after it}} }}  \,   {\Tilde {\bf a}}_{s,i_1}(l_{\tau_{s,i_1}}, \nu_{s,i_1})^H {\Tilde {\bf a}}_{s,i_2}(l_{\tau_{s,i_2}}, \nu_{s,i_2})  \nonumber \\
& &  \hspace{-23mm}  +  \,  E_p  \sum\limits_{s_1=1}^K  \sum\limits_{\substack{s_2=1  \\  s_2  \ne  s_1}}^K \sum\limits_{i_1=1}^{L_{s_1}} \sum\limits_{i_2=1}^{L_{s_2}}    \underbrace{{\Bigg (}  \frac{\sum\limits_{q=1}^Q   h_{q,s,i_1}^* h_{q,s,i_2}   }{Q}  {\Bigg )}}_{\substack{\rightarrow 0, \, \mbox{\small{as}}  \, (Q_h, Q_v) \rightarrow \infty  \,  ,\, \\ \mbox{\small{see}}  \,  (\ref{eqnvvp12} )  \, \mbox{\small{and para. after}} \, (\ref{eqnvvp34})  }}  \,   {\Tilde {\bf a}}_{s,i_1}(l_{\tau_{s,i_1}}, \nu_{s,i_1})^H {\Tilde {\bf a}}_{s,i_2}(l_{\tau_{s,i_2}}, \nu_{s,i_2}).
\end{eqnarray}
\normalsize}
\vspace{-3mm}
\end{figure*}
From (\ref{eqn2223}) we see that when the number of BS antennas is sufficiently large, a good approximation to the ML objective function is given by (\ref{eqn2324}) (see top of next page).
\begin{figure*}
{
\vspace{-5mm}
\small
\begin{eqnarray}
\label{eqn2324}
\sum_{q=1}^Q  \left\Vert  {\widehat {\bf x}_q}  \, -  \, \sqrt{E_p} \sum_{s=1}^K \sum\limits_{i=1}^{L_s} h_{q,s,i}  \, {\Tilde {\bf a}}_{s,i}(l_{\tau_{s,i}}, \nu_{s,i}) \right\Vert^2  & \hspace{-2mm}  \approx  &  \hspace{-2mm} \sum_{q=1}^Q  \Vert   {\widehat {\bf x}_q}  \Vert^2  \, - \, 2  \sqrt{E_p}  \sum\limits_{s=1}^K \sum\limits_{i=1}^{L_s} \sum\limits_{q=1}^Q \Re \left[ h_{q,s,i}  \,  {\widehat {\bf x}_q}^H  \, {\Tilde {\bf a}}_{s,i}(l_{\tau_{s,i}}, \nu_{s,i}) \right]  \nonumber \\
&  &   + \, E_p \sum\limits_{s=1}^K \sum\limits_{i=1}^{L_s}  \sum\limits_{q=1}^Q \vert h_{q,s,i} \vert^2   \,  \Vert  {\Tilde {\bf a}}_{s,i}(l_{\tau_{s,i}}, \nu_{s,i}) \Vert^2.
\end{eqnarray}
\vspace{-4mm}
\begin{eqnarray*}
\hline
\end{eqnarray*}
\normalsize}
\vspace{-4mm}
\end{figure*}
Using the large antenna approximation of (\ref{eqn2324}) in (\ref{eqn2123}), the ML estimates are approximately given by (\ref{eqn2425}) (see top of the page after next page).
Here we have also used the fact that $\Vert  {\Tilde {\bf a}}_{s,i}(l_{\tau_{s,i}}, \nu_{s,i}) \Vert^2 = 1$ (see (\ref{AsiHAsi})).
\begin{figure*}
{
\vspace{-7mm}
\small
\begin{eqnarray}
\label{eqn2425}
\left( \{ \widehat{h_{q,s,i}} \}_{q=1,s=1,i=1}^{Q,K, L_s}, \{ \widehat {l_{\tau_{s,i}}}   \}_{s=1,i=1}^{K, L_s} ,   \{ \widehat{\nu_{s,i}}   \}_{s=1,i=1}^{K, L_s}  \right) & \hspace{-2mm}  \approx &  \hspace{-2mm} \arg  \hspace{-9mm}  \max_{\substack{   \\  \\ \{ h_{q,s,i} \}_{q=1,s=1,i=1}^{Q,K, L_s} \\
\{ l_{\tau_{s,i}}   \}_{s=1,i=1}^{K, L_s} \\ \{ {\nu_{s,i}}   \}_{s=1,i=1}^{K, L_s}} }  \hspace{-6mm}  \sum\limits_{s=1}^K\sum\limits_{i=1}^{L_s} {\Bigg [}   \sum\limits_{q=1}^Q  {\Bigg (}  2 \sqrt{E_p}   \Re \left[ h_{q,s,i}  \,  {\widehat {\bf x}_q}^H  \, {\Tilde {\bf a}}_{s,i}(l_{\tau_{s,i}}, \nu_{s,i}) \right]    - \, E_p \,  \vert h_{q,s,i} \vert^2    {\Bigg )} \, {\Bigg ]}. \nonumber \\
\end{eqnarray}
\normalsize}
\vspace{-2mm}
\end{figure*}
Due to the separation of the terms for each UT and each path, in the R.H.S. of (\ref{eqn2425}), separate estimation of
each UT's channel path gains, path delays and Doppler shifts can be performed.
For the $i$-th path of the $s$-th UT, the proposed estimate of $l_{\tau_{s,i}}, \nu_{s,i}$ and $h_{q,s,i}, q=1,2,\cdots, Q$
is given by (\ref{eqn2526}) (see top of next page).
\begin{figure*}
{
\vspace{-11mm}
\small
\begin{eqnarray}
\label{eqn2526}
\left( \{ \widehat{\widehat{h_{q,s,i}}} \}_{q=1}^{Q},  \widehat{\widehat {l_{\tau_{s,i}}}},   \widehat{\widehat{\nu_{s,i}}}  \right) & \hspace{-2mm}  \Define &  \hspace{-2mm} \arg  \hspace{-3mm}  \max_{\substack{   \\ \{ h_{q,s,i} \}_{q=1}^{Q} \\
 l_{\tau_{s,i}}  \,  ,  \, {\nu_{s,i}}    } }  \,  {\Bigg [}   \sum\limits_{q=1}^Q  {\Bigg (}  2 \sqrt{E_p}   \Re \left[ h_{q,s,i}  \,  {\widehat {\bf x}_q}^H  \, {\Tilde {\bf a}}_{s,i}(l_{\tau_{s,i}}, \nu_{s,i}) \right]    - \, E_p \,  \vert h_{q,s,i} \vert^2    {\Bigg )} \, {\Bigg ]}. 
\end{eqnarray}
\normalsize}
\end{figure*}
For a given choice of $( l_{\tau_{s,i}}  \,  ,  \, {\nu_{s,i}})$, the choice of $h_{q,s,i}$ which maximizes the objective function in (\ref{eqn2526}) is given by

{\vspace{-4mm}
\small
\begin{eqnarray}
\label{heqsiest23}
\widehat{\widehat{h_{q,s,i}}}(l_{\tau_{s,i}}  \,  ,  \, {\nu_{s,i}}) & = & \frac{1}{\sqrt{E_p}} \, {\Tilde {\bf a}}_{s,i}(l_{\tau_{s,i}}, \nu_{s,i})^H {\widehat {\bf x}_q}.
\end{eqnarray}
\normalsize}
Using this in (\ref{eqn2526}) we further get (\ref{eqn2627}) (see top of next page).
\begin{figure*}
{\vspace{-9mm}
\small
\begin{eqnarray}
\label{eqn2627}
\left(  \widehat{\widehat {l_{\tau_{s,i}}}},   \widehat{\widehat{\nu_{s,i}}}  \right) & \hspace{-2mm}  \Define &  \hspace{-2mm} \arg  \hspace{-3mm}  \max_{\substack{   \\
 l_{\tau_{s,i}}  \,  ,  \, {\nu_{s,i}}    } }  \,  {\Bigg [}   \sum\limits_{q=1}^Q  {\Bigg (}  2 \sqrt{E_p}   \Re \left[  \widehat{\widehat{h_{q,s,i}}}(l_{\tau_{s,i}}  \,  ,  \, {\nu_{s,i}}) \,  {\widehat {\bf x}_q}^H  \, {\Tilde {\bf a}}_{s,i}(l_{\tau_{s,i}}, \nu_{s,i}) \right]    - \, E_p \,  \left\vert  \widehat{\widehat{h_{q,s,i}}}(l_{\tau_{s,i}}  \,  ,  \, {\nu_{s,i}})  \right\vert^2    {\Bigg )} \, {\Bigg ]}   \nonumber \\
& \hspace{-2mm}  =  &  \hspace{-2mm} \arg  \hspace{-3mm}  \max_{\substack{   \\  l_{\tau_{s,i}}  \,  ,  \, {\nu_{s,i}}    } }  \,   \sum\limits_{q=1}^Q  \left\vert   {\Tilde {\bf a}}_{s,i}(l_{\tau_{s,i}}, \nu_{s,i})^H {\widehat {\bf x}_q}    \right\vert^2.
\end{eqnarray}
\vspace{-4mm}
\begin{eqnarray*}
\hline
\end{eqnarray*}
\normalsize}
\vspace{-3mm}
\end{figure*}  
Due to the non-linear dependence of ${\Tilde {\bf a}}_{s,i}(l_{\tau_{s,i}}, \nu_{s,i})$ on $(l_{\tau_{s,i}}, \nu_{s,i})$, it is difficult to derive a closed form expression for
$\left(  \widehat{\widehat {l_{\tau_{s,i}}}},   \widehat{\widehat{\nu_{s,i}}}  \right)$. Therefore, we propose the following method.
Since $0 \leq {l_{\tau_{s,i}}} \leq (M-1)$ and ${l_{\tau_{s,i}}} \in {\mathbb Z}$, we firstly estimate the number of channel paths ${\widehat L_s}$ for the $s$-th UT and the delay ${l_{\tau_{s,i}}}, i=1,2,\cdots, {\widehat L_s}$ for each path of the $s$-th UT by observing the
distribution of the received energy in the pilot region ${\mathcal R}_s$ dedicated for the $s$-th UT. Towards this end, for any $l$ s.t. $(k,l) \in {\mathcal R}_s$ for some $k$, the energy received in the $l$-th DDRE along the delay domain is given by

{\vspace{-4mm}
\small
\begin{eqnarray}
\label{eqnesl}
{\mathcal E}_s[l] & \Define &  \frac{\sum\limits_{\substack{k  \, | \, (k,l) \in {\mathcal R}_s} }  \sum\limits_{q=1}^Q  \left\vert    {\widehat x_q}[k,l]  \right\vert^2  }{Q}.
\end{eqnarray}     
\normalsize}
The proposed estimate of the number of channel paths between the $s$-th UT and the BS is then given by the number of delay domain indices $l$ for which
${\mathcal E}_s[l]$ exceeds a threshold ${\mathcal E}_{th}$, i.e.
\begin{eqnarray}
\label{estLseqn}
{\widehat L_s} & = &  \sum\limits_{\substack{l  \, | \, (k,l) \in {\mathcal R}_s} \, \mbox{\tiny{for some}} \, k}  {{1}}_{_{{\mathcal E}_s[l] \, > \, {\mathcal E}_{th}}}
\end{eqnarray} 
where $1_{_{{\mathcal E}_s[l] \, > \, {\mathcal E}_{th}}}$ is one when ${\mathcal E}_s[l] \, > \, {\mathcal E}_{th}$ and is otherwise zero.
The ${\widehat L_s}$ estimated path delays $\widetilde{l_{\tau_{s,i}}}, i=1,2,\cdots, {\widehat L_s}$ are then given by those indices $l$ for which ${\mathcal E}_s[l] \, > \, {\mathcal E}_{th}$
and $l$ is such that $(k,l) \in {\mathcal R}_s$ for some $k$.
The threshold ${\mathcal E}_{th}$ is chosen such that false detection of channel paths which do not exist is extremely rare.
%so that the achieved error rate performance with the estimated channel is almost same as the error rate performance obtained with
%perfect knowledge of the channel path gains, delay and Doppler shifts.
For the simulations presented in Section \ref{errorrateperf},
we have considered ${\mathcal E}_{th} = 4 N N_o$, i.e., four times the mean value of ${\mathcal E}_s[l]$ in the presence of only AWGN (i.e., if no pilots are transmitted).   

Using (\ref{eqn2627}) the proposed estimate for the Doppler shift of the $i$-th detected path of the $s$-th UT is given by

{\vspace{-5mm}
\small
\begin{eqnarray}
\label{eqn2728}
\widetilde{\nu_{s,i}}  &  \Define   &    \hspace{-1mm} \arg  \hspace{-7mm}  \max_{\substack{   \\   - \nu_{max} \leq  {\nu_{s,i}}   \leq \nu_{max}  } }  \,   \sum\limits_{q=1}^Q  \left\vert   {\Tilde {\bf a}}_{s,i}(\widetilde{l_{\tau_{s,i}}}, \nu_{s,i})^H {\widehat {\bf x}_q}    \right\vert^2.
\end{eqnarray}
\normalsize}
In the absence of a closed-form expression for $\widetilde{\nu_{s,i}}$,
for the simulations presented in Section \ref{errorrateperf} we maximize the objective function in (\ref{eqn2728}) over
uniformly spaced $400$ values of ${\nu_{s,i}}$ in the interval $ [ - \nu_{max} \,,\,   \nu_{max} ]$.             
Finally, just as in (\ref{heqsiest23}), the proposed estimate of $h_{q,s,i}$ for all $q=1, \cdots, Q$, $s=1,2,\cdots, K$ and $i=1,2,\cdots, {\widehat L_s}$ is given by

{\vspace{-5mm}
\small
\begin{eqnarray}
\label{heqsiest28}
{\widetilde{h_{q,s,i}}} & = & \frac{1}{\sqrt{E_p}} \, {\Tilde {\bf a}}_{s,i}(\widetilde{l_{\tau_{s,i}}}, \widetilde{\nu_{s,i}})^H {\widehat {\bf x}_q}.
\end{eqnarray}
\normalsize
}
Just as in (\ref{eq_channel}), the proposed estimate of the DD domain channel between the $s$-th UT and the $q$-th BS antenna is

{\vspace{-5mm}
\small
\begin{eqnarray}
\label{eq_est_channel}
\widetilde{{\bf H}_{q,s}} \Define \sum\limits_{i=1}^{\widehat{L_s}}    {\widetilde{h_{q,s,i}}}    \,  \widetilde{{\bf A}_{s,i}} 
\end{eqnarray}
\normalsize}
where the element in the $(kM + l+1)$-th row and $(k'M + l' + 1)$-th column of $\widetilde{{\bf A}_{s,i}}  \in {\mathbb C}^{MN \times MN}$
is nothing but the expression for $a_{s,i,k,l}[k',l']$ in (\ref{asikleqn}) with $l_{\tau_{s,i}}$ and $\nu_{s,i}$ replaced by their proposed estimates $\widetilde{l_{\tau_{s,i}}}$ and $\widetilde{\nu_{s,i}}$ respectively.

From (\ref{eqnesl}), (\ref{estLseqn}), (\ref{eqn2728}), (\ref{heqsiest28}) and (\ref{eq_est_channel}), it follows that the complexity of computing the proposed channel matrix estimates $\widetilde{{\bf H}_{q,s}}, q=1,2,\cdots, Q$
for a particular UT is $O(Q{\widehat{L_s}}MN^2)$.
Here we have also used the fact that the vector ${\Tilde {\bf a}}_{s,i}(\widetilde{l_{\tau_{s,i}}}, \nu_{s,i})$ has only $N$ non-zero entries (see the expression for $a_{s,i,k,l}[\cdot, \cdot]$ in (\ref{asikleqn})).
As the DD domain channel matrix ${\bf H}_{q,s}$ has $MN^2$ non-zero entries, there are $QMN^2$ non-zero entries in all the $Q$ channel matrices ${\bf H}_{q,s}, q=1,2,\cdots, Q$ for a given UT. Therefore, the complexity of estimating each non-zero entry is only $O({\widehat{L_s}})$ (i.e., independent of $(Q,M,N)$).

\end{document}